\documentclass[notitlepage,12pt]{article}

\usepackage[paper=letterpaper,hmargin=2.5cm,top=2.3cm,bottom=2.3cm,bindingoffset=0cm,twoside,hoffset=0pt,marginparsep=0pt,marginparwidth=0pt]{geometry}

\usepackage{placeins} 
\usepackage{nicematrix}
\usepackage{amssymb}
\usepackage{amsmath}
\usepackage{tikz}
\usetikzlibrary{tikzmark}
\usepackage{amsfonts}
\usepackage{graphicx}
\usepackage{xcolor}
\usepackage{colortbl}
\usepackage{xspace}
\usepackage{ulem}
\usepackage{wasysym}
\usepackage[toc,page]{appendix}
\usepackage{cancel}
\usepackage{slashed}
\usepackage{anyfontsize}
\usepackage[11pt]{moresize}

\usepackage{cite}

\usepackage{multirow,rotating}

\def\gsim{\raise0.3ex\hbox{$\;>$\kern-0.75em\raise-1.1ex\hbox{$\sim\;$}}}
\def\lsim{\raise0.3ex\hbox{$\;<$\kern-0.75em\raise-1.1ex\hbox{$\sim\;$}}}

\newcommand{\ba}[1]{\begin{eqnarray} \label{(#1)}}
\newcommand{\ea}{\end{eqnarray}}

\def\gsim{\raise0.3ex\hbox{$\;>$\kern-0.75em\raise-1.1ex\hbox{$\sim\;$}}}
\def\lsim{\raise0.3ex\hbox{$\;<$\kern-0.75em\raise-1.1ex\hbox{$\sim\;$}}}

\title{Leading large $N_c$ contributions to Lepton Number Violating Meson Decays}

\author{
Andrea Donini$^1$\footnote{donini@ific.uv.es},
Marcela Gonz\'alez$^2$\footnote{marcela.gonzalezpi@uv.cl},
Martin Hirsch$^1$\footnote{mahirsch@ific.uv.es},
Nicol\'as A. Neill$^3$\footnote{nclsneill@gmail.com}
\\[2.5ex]
\small
$^1$\textit{Instituto de F\'{\i}sica Corpuscular --
  CSIC - Universitat de Val{\`e}ncia, Parc Cient\'{\i}fic-UV,}\\
\small
\textit{  c/ Catedr\'atico Jos\'e Beltr\'an, 2,
  E-46980 Paterna (Val{\`e}ncia), Spain}\\
\small
$^2$\textit{Instituto de F\'isica y Astronom\'ia, Universidad de Valpara\'iso,}\\
\small
\textit{Avenida Gran Breta\~na 1111, Valpara\'iso, Chile}\\
\small
$^3$\textit{Centro Multidisciplinario de F\'isica, Vicerrector\'ia de Investigaci\'on, Universidad Mayor,}\\
\small
\textit{8580745 Santiago, Chile}
}

\date{}

\begin{document}

\maketitle


\begin{abstract}
Lepton number violating meson decays, such as $M_1^- \to
M_2^+\ell_1^-\ell_2^-$, provide
constraints on $d=9$ $\Delta L = 2$ operators. RGE-improved bounds on
the Wilson coefficients of these operators have been presented in the literature, taking into account perturbative QCD one-loop corrections and the corresponding operator mixing. Here, we present 
for the first time the contribution of connected diagrams to the hadronic matrix elements
$\langle M_2 | {\cal O}_h | M_1 \rangle$. These diagrams, usually
overlooked under the assumption that $\langle M_2 | {\cal O}_h | M_1
\rangle \sim \langle M_2 | J_{q_3 q_4} | 0 \rangle \times \langle 0 |
J_{q_1 q_2} | M_1 \rangle \gg \langle M_2 | J_{q_3 q_2} \times J_{q_1
  q_4} | M_1 \rangle$, can give indeed a significant contribution to
the matrix element. Including these connected diagrams is but the
first step towards a full non-perturbative computation of the
long-range QCD effects in these operators, that should be performed
using lattice field theory techniques. However, connected diagrams
represent the leading order in the $1/N_c$ expansion of the QCD
non-perturbative effects and thus our work can be understood as
a realistic, first approximation to a complete calculation of the
long-range part of the matrix elements.
\end{abstract}

\newpage
\section{Introduction}
\label{sec:introduction}

In the standard model lepton number is conserved at the perturbative
level. However, in Standard Model Effective Field Theory (SMEFT)
\cite{Grzadkowski:2010es,Henning:2014wua,deBlas:2017xtg,Falkowski:2017pss,Brivio:2017vri} lepton number violation (LNV) appears in the form of
non-renormalizable operators \cite{Weinberg:1979sa,Babu:2001ex}. The
most stringent test of lepton number violation is neutrinoless double
beta decay ($0\nu\beta\beta$), for reviews see for example
\cite{Rodejohann:2011mu,Deppisch:2012nb, Dolinski:2019nrj}.

However, $0\nu\beta\beta$ is sensitive only to LNV in the first
generation of leptons. To test LNV for higher generation leptons one
has to resort to exotic processes, such as $\tau^{\pm} \to \ell^{\mp}
M_1 M_2$, and $M_1^- \to M_2^+\ell_1^-\ell_2^-$ where $\ell$ stands
for $\ell=e, \mu$ in case of tau decays and, in principle, any SM
generation in meson decays (when kinematically allowed).  
Experimental constraints for all of these
processes can be found in the Particle Data Book
\cite{ParticleDataGroup:2024cfk}, where the results from flavour
factories~\cite{NA62:2022tte,NA62:2019eax,BaBar:2011ouc,BaBar:2012eip,
  NA62:2021zxl,BELLE:2011bej} and flavoured experiments at the
LHC~\cite{LHCb:2020car,LHCb:2014osd,LHCb:2012pcm,LHCb:2011yaj} are
summarized.

Comparing the experimental data with the theoretical predictions in
the lepton number violating sector of the SMEFT allows to put
constraints on the Wilson coefficients of individual effective
operators, for earlier work considering constraints from neutrino
masses and/or LNV meson decays see \cite{Littenberg:1991ek,
  Ilakovac:1995wc,deGouvea:2007qla, Helo:2010cw,Quintero:2011yh,
  LopezCastro:2012udb,Dong:2013raa, Quintero:2016iwi,Yuan:2017xdp,
  Mandal:2017tab,Cvetic:2017vwl, Yuan:2017uyq,Abada:2017jjx,
  Rendon:2019awg,Chun:2019nwi,Liao:2019gex,Liao:2020roy,
  Deppisch:2020oyx,Liao:2021qfj}, while results for LNV muon
conversion can be found, for example, in \cite{Simkovic:2000ma,
  Simkovic:2001fs,Berryman:2016slh,Yeo:2017fej}.\footnote{We note in
passing that also searches at LHC can put bounds on LNV models
\cite{Helo:2013dla,Helo:2013ika,Peng:2015haa,Gonzalez:2016ztm}.}
However, the matching between the experimental data and the
theoretical prediction is affected by a number of uncertainties
related to the perturbative order at which the theoretical computation
of the Wilson coefficients has been performed and on the specific
regularization and renormalization schemes adopted. The running of the
Wilson coefficients between the scale at which the theoretical
(perturbative) computation of short-range effects is sound and the
(lower) scale at which the experimental measurements are made must be taken
into account, introducing scheme dependence. Moreover, quite generally
it is found that loop computations introduce mixing of the operators,
responsible for a given transition at tree-level, with other operators
originally not present, thus enlarging the operator basis to be taken
into account. This was shown to be also the case for $M_1^- \to
M_2^+\ell_1^-\ell_2^-$ and $\tau^- \to \ell^+ M_1^- M_2^-$ decays
\cite{Gonzalez:2023him}, where QCD one-loop corrections where computed
and Renormalization Group (RG) running was properly accounted for.
The inclusion of these effects was shown to have a significant impact
on the bounds on the relevant Wilson coefficients.

In this paper, our aim is to go one step beyond the work done in
\cite{Gonzalez:2023him}, and introduce a first estimate of the effect
of long-range physics. This can be done by computing the relevant
hadronic matrix element at leading order in the $1/N_c$ expansion.

The hadronic matrix element $\langle M_2 | {\cal O}_h | M_1 \rangle$
is usually computed under the assumption that the s-channel--like
contribution dominates, {\it i.e.} that $\langle M_2 | {\cal O}_h |
M_1 \rangle \simeq \langle M_2 | J_{q_3 q_4} \times J_{q_1 q_2} | M_1
\rangle$, with $J_{qq}$ a quark current and where the initial and
final mesons sources are $M_1 = \bar q_1 \Gamma q_2$ and $M_2 = \bar
q_3 \Gamma^\prime q_4$, with $\Gamma, \Gamma^\prime$ some
$\gamma$-matrix dependent on the mesons quantum numbers. In order to
get an estimate of the hadronic matrix element, the {\it vacuum
  insertion approximation} (VIA) is usually adopted: $\langle M_2 |
J_{q_3 q_4} \times J_{q_1 q_2} | M_1 \rangle \sim \langle M_2 | J_{q_3
  q_4} |0 \rangle \, \langle 0 | J_{q_1 q_2} | M_1 \rangle $. This
approximation corresponds to neglecting all soft gluon
exchanges between the initial and final mesons and, at the same
time, to neglecting the production of intermediate hadronic states (such
as, for example, other heavier mesons with the appropriate quantum
numbers). In the language of the $1/N_c$ expansion, the resulting {\it
  disconnected} diagram is ${\cal O} (N_c^2)$, as it contains two
fermion loops, contributing $N_c$ each. All other possible diagrams,
on the other hand, are sub-leading under $N_c$--power-counting.  We
will, therefore, compute here the leading $N_c$ contribution to the
hadronic matrix element, represented by the {\it connected} diagram
$\langle M_2 | J_{q_3 q_2} \times J_{q_1 q_4} | M_1 \rangle$ in which
the quarks in ${\cal O}_h$ connect to external quark lines with 
t-channel--like contractions.  We will demonstrate how including this
correction modifies the bounds on the Wilson coefficients, in some
cases by considerable factors. 

The rest of this paper is organized as follows: in
Sect.~\ref{sec:operatorbasis} we consider the case of the $K^+ \to
\pi^- \, \ell^+ \, \ell^+$ decay as an example and present the
low-energy effective operators basis responsible for this process; 
in Sect.~\ref{sec:perturbative} we compute the decay widths 
and run the Wilson coefficients down at the hadronic scale $\mu$; 
in Sect.~\ref{sec:disconnected} we introduce the {\it vacuum insertion
approximation} (VIA) and compute the leading contribution to the hadronic matrix element in $1/N_c$; in
Sect.~\ref{sec:connected} we compute the ${\cal O } (1/N_c)$ correction
to the hadronic matrix element; in Sect.~\ref{sec:results} we derive
improved bounds to the Wilson coefficients at the hadronic scale $\mu$
at the sub-leading order in $1/N_c$; finally, in Sect.~\ref{sec:concl} we present our conclusions. 
In App.~\ref{app:ADM} we summarize the anomalous dimensions at one-loop relevant for the process at hand; 
in App.~\ref{app:umatrices}
we show the numerical values of the RGE evolution matrices;
in App.~\ref{app:fierz} we list the Fierz identities used in the paper;
in App.~\ref{app:sums} we give the sum of connected and disconnected diagrmas in the VIA.
Finally, in App.~\ref{app:Bparameters}, we introduce an optimal definition of the $B$-parameters to be computed on the lattice in order to perform a numerical computation of the hadronic matrix elements.

\section{Dimension-9 operators for $\Delta L = 2$ transitions}
\label{sec:operatorbasis}

\begin{figure}
	\centering
    \begin{tabular}{cc}
	\includegraphics[scale=0.50]{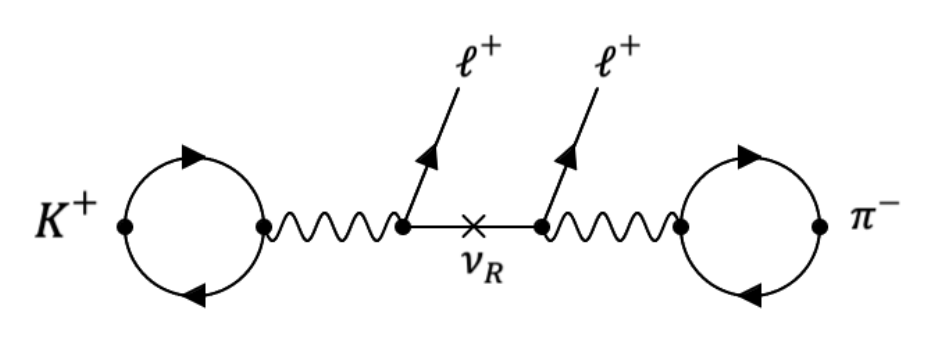} &
	\includegraphics[scale=0.40]{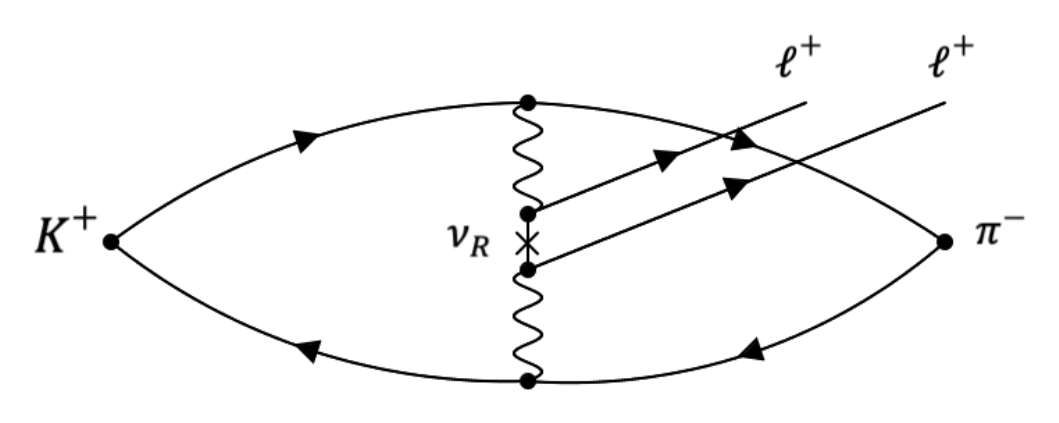}
	\end{tabular}
    \caption{\it Tree-level diagrams responsible for the $K^+ \to \pi^- \, \ell^+ \ell^+$ transition in the type-I see-saw model. 
    Left: s-channel amplitude. Right: t-channel amplitude.
    }
    \label{fig:seesawdiagrams}
\end{figure}

The process $K^+ \to \pi^- \ell^+ \ell^+ $ is the paradigmatic $\Delta L = 2$ meson decay. It can occur if an operator that violates lepton number by 2 units exists, such as for example in type-I see-saw models where a heavy Majorana ``right-handed" neutrino is added to the Standard Model spectrum. 

\subsection{$K^+ \to \pi^- \ell^+ \ell^+$ in type-I see-saw models}

Within this particular model, that we illustrate as an example, it was shown in Refs.~\cite{Littenberg:1991ek,Littenberg:2000fg} 
that this decay may occur through two tree-level diagrams 
(see Fig.~\ref{fig:seesawdiagrams}): 
\begin{enumerate}
\item {\bf s-channel}, Fig.~\ref{fig:seesawdiagrams} (left): in this case, the process goes as $K^+ \to W^+ \to \ell^+ \ell^+ W^- \to \pi^-$. 
Assuming that the $\Delta L = 2$ operator arises through the exchange of $N_R$ heavy Majorana neutrinos of mass $M_{\nu_j}$, we have: 
\begin{equation}
L^{\mu\nu}_j (p,p^\prime) = M_{\nu_j} \frac{1}{q^2 - M_{\nu_j}^2} \bar v (p) \gamma^\mu \gamma^\nu P_R v^c (p^\prime) \, .
\end{equation}
Eventually, the amplitude for the s-channel diagram is (after integrating out the W's):
\begin{equation}
\label{eq:schannelseesaw}
A_s = 2 G_F^2 f_K f_\pi \left [ V_{ud} V_{us} \right ]^\star \sum_{j = 1}^{N_R} \left [ U_{\ell j} U_{\ell^\prime j} \right ]^\star \, 
p_K^\mu p_\pi^\nu \, \left [ L_{\mu\nu,j} (p,p^\prime) - \delta_{\ell \ell^\prime} L_{\mu\nu, j} (p^\prime,p) \right ],
\end{equation}
where the decay constants $f_K,f_\pi$ come by neglecting soft gluons interchange between initial and final mesons and
applying the PCAC to the hadronic matrix elements $ \langle 0|\gamma_\mu \gamma_5 |K (q) \rangle$
and $\langle 0|\gamma_\mu \gamma_5 | \pi (q-p-p^\prime) \rangle$. 
\item {\bf t-channel}, Fig.~\ref{fig:seesawdiagrams} (right): in this case, the $W$'s are produced from two different kaon quark lines and, therefore, 
the hadronic matrix element is more complicated as, 
strictly speaking, it is a four-point function: $\langle \pi (z) | \bar u (y) \gamma^\mu_L d (y) \bar u (x) \gamma^\nu_L s(x) | K (0) \rangle$. In Fourier transform:
\begin{eqnarray}
A_t &=& 2 G_F^2  \left [ V_{ud} V_{us} \right ]^\star \sum_{j = 1}^{N_R} \left [ U_{lj} U_{l^\prime j} \right ]^\star \, 
 \left [ L_{\mu\nu,j} (p,p^\prime) - \delta_{l l^\prime} L_{\mu\nu, j} (p^\prime,p) \right ] \nonumber \\
& \otimes & 
\int \, d^4 x \, d^4 y \, e^{i (p_d - p_u) y} \, e^{i(p_{\bar s} - p_{\bar u}) x} \langle \pi (z) | \bar u (y) \gamma^\mu_L d (y) \bar u (x) \gamma^\nu_L s(x) | K (0) \rangle \, .
\end{eqnarray}
If $q^2 \ll M_{\nu_j}$ for any $j$, we can integrate out the heavy right-handed neutrinos and get a local lepton current:
\begin{equation}
L^{\mu\nu}_j (p,p^\prime) \sim - \frac{1}{m_{\nu_j}} \bar v (p) \gamma^\mu \gamma^\nu P_R v^c (p^\prime) + \dots
\end{equation}
After integrating out both $W$'s and the heavy right-handed neutrinos, the two quark currents give a local four-fermion $\Delta Q = 2$ hadronic
current-current operator ${\cal O}_h = J_h \times J_h$ (with $J_h = \bar q_i \Gamma q_j$ a fermion bilinear), such that the hadronic amplitude is:
\begin{equation}
{\cal M} (z,x) = \langle \pi (z) | {\cal O}^{\Delta Q = 2,LL}_{\mu\nu} | K (0) \rangle = \langle \pi (z) | \bar u (x) \gamma_\mu^L d (x) \bar u (x) \gamma_\nu^L s(x) | K (0) \rangle\, .
\end{equation} 
This matrix element is a three-point function and it must be computed non-perturbatively, for example on the lattice. 
\end{enumerate}

\begin{figure}
	\centering
    \begin{tabular}{cc}
	\includegraphics[scale=0.50]{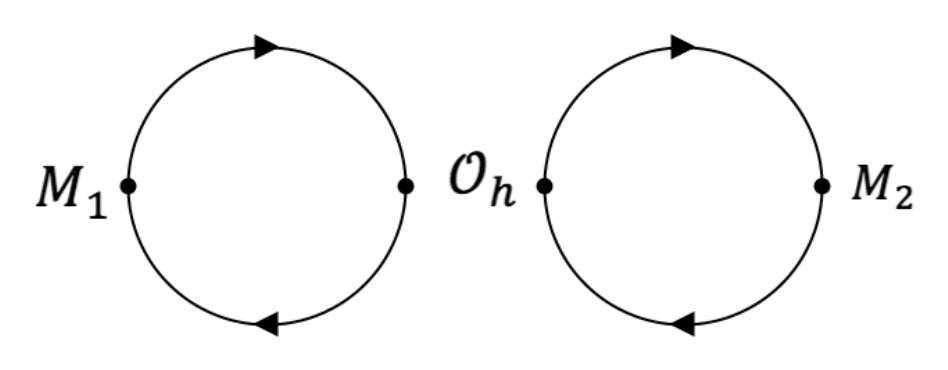} &
	\includegraphics[scale=0.50]{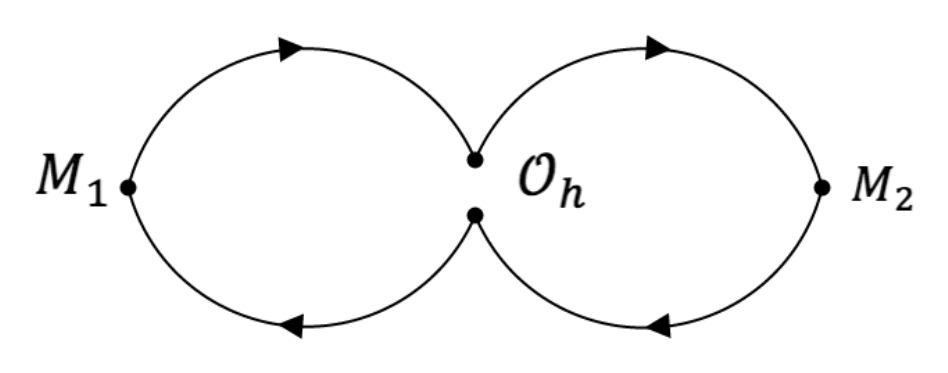}
	\end{tabular}
    \caption{\it Disconnected (left) and connected (right) diagrams of the hadronic matrix element for the decay $M^+_1 \to M^-_2 \ell^+ \ell^+$, with
    effective operator ${\cal O} = {\cal O}_h \times J_{\ell \ell}$. Left diagram is ${\cal O}(N_c^2)$ and right diagram is ${\cal O}(N_c)$, as they contain two 
    and one fermion loops, respectively.
    }
    \label{fig:treelevel}
\end{figure}

As we have seen, if we are able to start from a definite fundamental theory, we have some hints on the spinor structure of the effective operators to be computed. In the case
of the type-I see-saw, we have an s-channel--induced product of two two-point functions (whose non-perturbative estimate may be obtained using the PCAC), or a t-channel--induced three-point function (for which we have no estimate, in principle). However, these two amplitudes are a quite general result independently of the original fundamental theory. 
Assume that we have a four-fermion local operator in the form  ${\cal O}_h = J_h \times J_h$ responsible for a $\Delta L = 2$ transition in combination with a leptonic current, 
${\cal O}_{\Delta L = 2} = {\cal O}_h \times J_{\ell \ell}$. When computing the hadronic contribution to the process, we must sandwich the local four-quark operator between hadronic states to get a (non-perturbative) hadronic matrix element. Once contracting over the external states the hadronic local operator ${\cal O}_h$, 
in general two contributions arise from the ${\cal T}$-product of the quark fields: one connected "eight-like" diagram (Fig.~\ref{fig:treelevel}, right), 
or two disconnected loops (Fig.~\ref{fig:treelevel}, left). It is easy to relate the two diagrams with the two different amplitudes in the type-I see-saw model: 
the disconnected diagram arises integrating out heavy fields in the s-channel, whereas the connected diagram arises from the t-channel. 
From a bottom-up approach such as the one used in SMEFT \cite{Grzadkowski:2010es,Henning:2014wua,deBlas:2017xtg,Falkowski:2017pss,Brivio:2017vri}, however, we will not be able to determine the specific structure 
of the diagrams in the fundamental theory, as we have in principle no idea of what the fundamental theory is. 

\subsection{$K^+ \to \pi^- \ell^+ \ell^+$ in the OPE}

If we have no fundamental theory to derive the low-energy operators relevant for the process $K^+ \to \pi^- \ell^+ \ell^+$, we can still write a complete basis of operators that may
be responsible for the transition, and then use the experimental data to constrain the corresponding Wilson coefficients. This is the approach followed in 
Refs.~\cite{Gonzalez:2015ady,Arbelaez:2016zlt,Arbelaez:2016uto,Gonzalez:2017mcg,Gonzalez:2023him}. 

The effective Lagrangian for a particular process that
violates $\Delta L$ by two units is: 
\begin{equation}
\mathcal L_{\rm eff}^{\Delta L=2} = \frac{1}{\Lambda_{\rm EW}^5} \, \sum_i C_i (\Lambda) 
\, {\cal O}_i^{\Delta L = 2}(\Lambda),
\label{eq:lagrangianLambda}
\end{equation}
where $C_i (\Lambda)$ is the {\it tree-level} 
Wilson coefficients of the {\it bare} operator ${\cal O}_i^{\Delta L = 2}$ and $i$ runs over the operator basis. We have normalized the Wilson 
coefficients with a common factor $1/\Lambda_{\rm EW}^5$, in order to obtain\footnote{The normalization scale is, of course, arbitrary. We find it convenient to use the scale of the electroweak symmetry breaking, $\Lambda_{\rm EW} = 246$ GeV, to compare directly with the case of the type-I see-saw model, in which the heavy mediators are $W$ bosons.} adimensional coefficients.
If we know the fundamental theory, the Wilson coefficients can be computed directly by matching the theory where heavy d.o.f.'s have been integrated out with the fundamental theory at some high-energy
scale $\Lambda$ (as it is done in the case of the $\Delta S = 2$ transitions in the SM, see Ref.~\cite{Buras:1998raa} and refs. therein). The coefficients $C_i$ can be then constrained
by experiments that look for a given process. 
If we do not know the fundamental theory, we can write down all possible operators that mediate the process at hand compatible
with the symmetries of the SM. 

For the process $K^+ \to \pi^- \ell^+_1 \ell^+_2$, it was found that the most general effective Lagrangian contains 7 operators in the
case $\ell_1 = \ell_2$, each of them with a ${\cal O}_h \times J_{\ell \ell}$ structure:
\begin{equation}
\label{eq:relevantoperators}
\left \{ 
\begin{array}{l}
{\cal O}_1^{XYZ} = 8 \left ( \bar u P_X d \, \bar u P_Y s \right ) \otimes j_{\ell \ell}^Z,  \\
\\
\mathcal{O}^{XXX}_{2}= 8 ({\bar u}\sigma^{\mu\nu}P_{X}d \, {\bar u}\sigma_{\mu\nu}P_{X}s) \otimes j^X_{\ell \ell}, 
\\
\\
{\cal O}_3^{XYZ} = 8 \left ( \bar u \gamma^\mu P_X d \, \bar u \gamma_\mu P_Y s \right ) \otimes j_{\ell \ell}^Z,  \\
\\
\mathcal{O}^{XYZ}_{4}= 8 ({\bar u}\gamma^{\mu}P_{X}d \, {\bar u}\sigma_{\mu}^{\,\,\nu}P_{Y}s) \otimes  j_{\ell \ell,\nu}^Z,
\\
\\
\mathcal{O}^{\prime \ XYZ}_{4}= 8 ({\bar u}\sigma_{\mu}^{\,\,\nu}P_X d  \, {\bar u}\gamma^{\mu}P_Y s)  \otimes j_{\ell \ell, \nu}^Z,
\\
\\
{\cal O}_5^{XYZ} = 8 \left ( \bar u \gamma^\mu P_X d \, \bar u P_Y s \right ) \otimes j^Z_{\ell \ell, \mu},  \\
\\
{\cal O}_5^{\prime \, XYZ} = 8 \left ( \bar u P_X d \, \bar u \gamma^\mu P_Y s \right ) \otimes j^Z_{\ell \ell, \mu},  \\
\end{array}
\right .
\end{equation}
where in these expressions $P_X$ are the chirality projectors, $P_{\rm L,R} =1/2 (1 \mp \gamma_5)$. 
The lepton bilinears $j$ for two generic leptons $\ell_1$ and $\ell_2$ are:
\begin{equation}
\left \{
\begin{array}{l}
j_{\ell_1 \ell_2}^Z = \bar \ell_1 \, P_Z \, \ell_2^c \, , \\
\\ 
j^Z_{\ell_1 \ell_2, \mu} = \bar \ell_1 \, \gamma_\mu \, P_Z \, \ell_2^c\, , \\
\\
j^Z_{\ell_1 \ell_2, \mu\nu} = \bar \ell_1 \, \sigma_{\mu\nu} \, P_Z \,  \ell_2^c \, .
\end{array} \label{eq:relevantoperators1b}
\right . 
\end{equation}

Notice that the tensor lepton current $j^Z_{\ell_1 \ell_2 \mu\nu}$ vanishes when $\ell_1=\ell_2$. This means that, when $\ell_1 \neq \ell_2$ (differently from 
the example that we were studying above, 
$K^+ \to \pi^- \ell^+ \ell^+$), new
operators should be added to our basis. 
Eventually, for $\ell_1 \neq \ell_2$, we should include four new operators: 
 \begin{equation}
\label{eq:relevantoperators2}
\left \{ 
\begin{array}{l}
\mathcal{O}^{XYZ}_{6} = 8 \, ({\bar u}\gamma^\mu P_{X}d \, {\bar u}\gamma^{\nu}P_{Y}s) \otimes j^Z_{\ell_1 \ell_2, \mu\nu} \, , \\
\\
\mathcal{O}^{XZZ}_{7} = 8 \, ({\bar u}P_{X}d \, {\bar u}\sigma^{\mu\nu}P_{Z} s) \otimes j^Z_{\ell_1 \ell_2, \mu\nu} \, , \\
\\
\mathcal{O}^{\prime ZXZ}_{7} = 8 \, ({\bar u}\sigma^{\mu\nu}P_{Z} d \, {\bar u}P_{X}s)  \otimes j^Z_{\ell_1 \ell_2, \mu\nu} \, ,\\
\\
\mathcal{O}^{XXX}_{8} = 8 \, ({\bar u }\sigma^{\mu\alpha} P_{X}d  \, {\bar u}\sigma^{\nu}_{\,\alpha}P_{X}s) \otimes  j^X_{\ell_1 \ell_2, \mu\nu} \, . 
\end{array}
\right .
\end{equation}
The primed operators $\mathcal{O}^{\prime XYZ}_{6}$ and $\mathcal{O}^{\prime XXX}_{8}$ are not included, 
since $\mathcal{O}^{\prime XYZ}_{6} = -\mathcal{O}^{YXZ}_{6}$ and $\mathcal{O}^{\prime XXX}_{8} = -\mathcal{O}^{XXX}_{8}$. Notice
that only particular combinations of chiral projectors are allowed for these additional operators.

Once we have introduced the full basis that may be responsible for the transition $K^+ \to \pi^- \ell_1^+ \ell_2^+$, it is easy to see that the operator that we have found in the case of the type-I see-saw is, in fact, ${\cal O}_3^{XXZ}$ for two identical final leptons, whereas
for processes with leptons from two different generations there will also be a contribution from  ${\cal O}_6^{XXZ}$.

\section{From the effective Lagrangian ${\cal L}_{\rm eff}$ to bounds on $C_i$}
\label{sec:perturbative}

The effective Lagrangian in eq.~(\ref{eq:lagrangianLambda}) must
be computed within definite initial and final states, in order to 
use the experimental data to constrain the Wilson coefficients. 
However, this must be done at some hadronic scale $\mu$ much lower
than the scale $\Lambda$ where the fundamental theory is replaced
by the effective one.

\subsection{Hadronic decay widths}
\label{sec:decaywidths}

The tree-level amplitudes for the process $K^+ \to \pi^- \ell^+ \ell^+$ mediated by the effective Lagrangian in Eq.\,(\ref{eq:lagrangianLambda}) are given by:
\begin{align}
    \mathcal M(K^+ \to \pi^- \ell_1^+ \ell_2^+) = \left<\pi^- \ell_1^+ \ell_2^+|\mathcal L_{eff}^{\Delta L=2}|K^+ \right>
    =
        \frac{1}{\Lambda_{\rm EW}^5} \, 
    \sum_n \sum_{\substack{XYZ}} C_{n}^{XYZ} (\Lambda) \, \mathcal A_{n}^{XYZ} (\Lambda),
\end{align}
where $\mathcal A_{n}^{XYZ}(\Lambda) = \left \langle \pi^- \ell_1^+ \ell_2^+|\mathcal O_{n}^{XYZ}(\Lambda)|K^+ \right \rangle =  \left \langle \pi^-|\mathcal O_{n}^{XYZ}(\Lambda)|K^+ \right \rangle \, \otimes J_{\ell_1 \ell_2}$. 
The amplitude ${\cal A}^{XYZ}_n$ depends on the scale $\Lambda$ since
the hadronic matrix elements are dressed by ${\cal O}(\Lambda)$ gluons.
The partial decay widths for the meson decays are then 
given by \cite{Quintero:2016iwi}:
\begin{align}
    \Gamma(K^+\to \pi^- \ell_1^+ \ell_2^+) = &
    \frac{1}{64 \, (2\pi)^3 \, m_{K}^3 \, \Lambda_{\rm EW}^{10}} \, 
 \int_{s^-}^{s^+} ds \int_{t^-}^{t^+} dt
    \left|
    \sum_n  \sum_{\substack{XYZ}} C_{n}^{XYZ} 
\mathcal{\overline{A}}_{n}^{XYZ}\right|^2,
\label{eq:mesondecaywidthtreelevel}
\end{align}
where $\mathcal{\overline{A}}_n$ are spin-averaged amplitudes 
and the kinematical variables  
$s\equiv (p_{\ell_1}+p_{\ell_2})^2$ and $t\equiv (p_{\ell_2}+p_{M_2})^2$, 
have integration limits given by:
\begin{align}
    s^{-} & = (m_{\ell_1}+m_{\ell_2})^2,\ \ \ s^{+} = (m_{M_1}-m_{M_2})^2,\\
    t^{\pm} & = m_{M_1}^2 + m_{\ell_1}^2 - \frac{1}{2s}\left[
    \left(s+m_{M_1}^2-m_{M_2}^2\right)
    \left(s+m_{M_{\ell_1}}^2-m_{M_{\ell_2}}^2\right)\right.\nonumber\\
    &
    \left.
    \ \ \ \ \ \
    \mp \, \lambda^{1/2}(s,m_{\ell_1}^2,m_{\ell_2}^2) \, 
    \lambda^{1/2}(s,m_{M_1}^2,m_{M_2}^2)
    \right],
\end{align}
with $\lambda(x,y,z)=x^2+y^2+z^2-2(xy+xz+yz)$. The quark flavor indices ($i,j,k,l$) are fixed by the quark content of the mesons: 
$M_1^-=(\bar u_i d_j)$, $M_2^+=(u_k \bar d_l)$.

The hadronic matrix element $\langle \pi^- | {\cal O}_n^{XYZ} (\Lambda)| K^+ \rangle$ should be in principle computed {\it non-perturbatively} at the same scale $\Lambda$ where the matching between the fundamental theory and the effective theory takes place. However, it is not possible to perform a non-perturbative QCD computation at that scale. What is instead done, is to compute
the matrix elements non-perturbatively at some low-energy scale 
$\mu \ll \Lambda$ where dynamical d.o.f.'s are mesons and baryons ({\it i.e.} at a scale where quarks are confined and only long-distance soft-gluons should be taken into account).

\subsection{One-loop QCD running of the Wilson Coefficients}
\label{sec:onelooprunning}

\begin{figure}[t]
\centering
\includegraphics[width=0.3\linewidth]{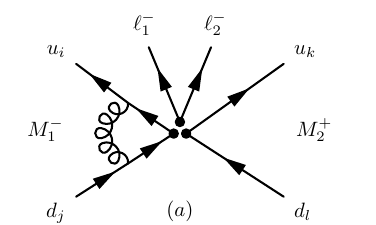}
\includegraphics[width=0.3\linewidth]{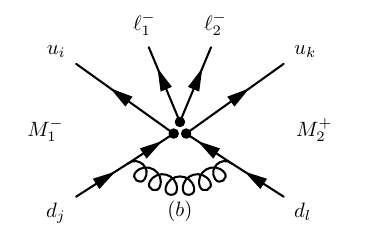}
\includegraphics[width=0.3\linewidth]{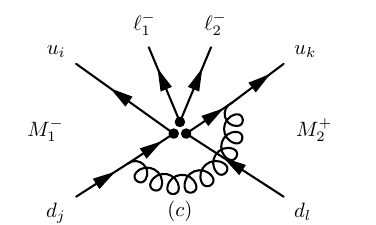}
\caption{\it Effective $d=9$ operator description of the short-range
  mechanisms (SRM) of the meson decay $M_1^{-} \to M_2^{+} \ell_1^- \ell_2^-$. The tree-level diagram can be obtained,
  e.g., by integrating out gauge bosons and heavy fermions
  in Fig.~\ref{fig:seesawdiagrams}. On the other hand, diagrams (a)-(c) give some representative one-loop perturbative QCD corrections to the
amplitude arising by the interchange of a hard gluon.}\label{fig:shortM}
\end{figure}

The main goal of the RGE running of the Wilson coefficients
is to connect the high-energy scale $\Lambda$, at which  the matching between the fundamental theory and the effective one is supposed to be performed, with some low-energy scale $\mu$ where experimental data are measured. The high-energy scale is typically 
$\Lambda = {\cal O}(100)$ GeV in LEFT, {\it i.e.} a scale large enough for
perturbative QCD to be reliable. On the other hand, the low-energy scale $\mu$ should be low enough for non-perturbative computations of the hadronic matrix element $\langle M_2 | {\cal O}^{\Delta L = 2}_i (\mu)| M_1 \rangle $ to be carried out. Depending on the particular non-perturbative approach used, this scale ranges\footnote{ 
The higher the low-energy scale, the more we can trust the perturbative
running of the Wilson coefficient. In lattice computation, 
typically $\mu = 2$ GeV. For other NP method, sometimes $\mu = 1$ GeV
is also chosen. Quite in general, though, to reach lower scales 
one-loop corrections are not enough and higher-order perturbative computations are needed \cite{Buras:2000if}.}
from 1 to 2 GeV.  The RGE running of the Wilson coefficients gives: 
\begin{equation}
\mathcal L_{\rm eff}^{\Delta L=2} = 
\frac{1}{\Lambda_{\rm EW}^5} \, 
\sum_i C_i (\mu) \, {\cal O}_i^{\Delta L = 2} (\mu)\, ,
\label{eq:lagrangian}
\end{equation} 
where the RG evolution, in a given regularization and renormalization
scheme, is given by the solution of the following equation:
\begin{equation}
\label{eq:RGE1}
\frac{d C_i(\mu)}{d \ln (\mu/\Lambda)} = \gamma_{ij} (\alpha_s) \, C_j(\mu),
\end{equation}
where $\gamma_{ij}$ is an element of the anomalous dimension matrix, $\hat{\gamma}$. The anomalous dimensions
of the operators in eqs.~(\ref{eq:relevantoperators}) and (\ref{eq:relevantoperators2}) have been computed in dimensional regularization in the $\overline{\mathrm{MS}}$ renormalization scheme 
in Ref.~\cite{Gonzalez:2023him}.  At leading order in $\alpha_s$, the anomalous dimension matrix takes the form: 
\begin{equation}
\label{eq:gamma}
\gamma_{ij}(\alpha_s) = \frac{\alpha_s}{4\pi} \gamma_{ij}, \quad \text{with} \quad \gamma_{ij} = -2 (b_{ij} - 2 C_F \delta_{ij}),
\end{equation}
where $C_F = (N_c^2 - 1)/(2 N_c)$ and $b_{ij}$ are determined 
from the one-loop QCD corrections. The perturbative results
of Ref.~\cite{Gonzalez:2023him} are summarized in App.~\ref{app:ADM}.
In order to give a feeling of the one-loop computations to be performed, some representative one-loop QCD diagrams contributing to the renormalization of the effective operators in the basis defined above are shown in Fig.~\ref{fig:shortM} (additional diagrams obtained via left-right or up-down reflections have also been taken into account). 

The solution to the RGE in Eq.~\eqref{eq:RGE1} can be expressed in terms of an evolution matrix $\hat{U}(\mu, \Lambda)$ connecting a high-energy scale $\Lambda$ to a lower scale $\mu$:
\begin{equation}
\label{eq:RGESolution}
C_i(\mu) = U_{ij}(\mu, \Lambda) \, C_j(\Lambda).
\end{equation}

At leading logarithmic accuracy, this evolution matrix is given by
\begin{equation}
\label{eq:UMatrix}
\hat{U}(\mu, \Lambda) = \hat{V} \, 
\left[\left( \frac{\alpha_s(\Lambda)}{\alpha_s(\mu)} \right)^{\hat \gamma_{\rm diag}/(2\beta_0)} \right] \, \hat{V}^{-1},
\end{equation}
where $\hat{V}$ diagonalizes the matrix $\hat{\gamma}$:
\begin{equation}
\label{eq:VMatrix}
\hat \gamma_{\rm diag} = \hat{V}^{-1} \hat{\gamma} \hat{V} \, .
\end{equation}

The RGE-improved amplitudes for the process 
$K^+ \to \pi^- \ell_1^+ \ell_2^+$ are:
\begin{align}
    \mathcal M(K^+ \to \pi^- \ell_1^+ \ell_2^+) = 
    \frac{1}{\Lambda_{\rm EW}^5} \, 
    \sum_n \sum_{\substack{XYZ}} \, 
    C_{n}^{XYZ}(\mu)
    \, \mathcal A_{n}^{XYZ} (\mu) \, ,
\end{align}
and the corresponding partial decay widths are, thus:
\begin{align}
    \Gamma(K^+\to \pi^- \ell_1^+ \ell_2^+) = &
    \frac{1}{64 \, (2\pi)^3 \, m_{K}^3 \, \Lambda_{\rm EW}^{10}} \, 
\int_{s^-}^{s^+} ds \int_{t^-}^{t^+} dt
    \left|
    \sum_n \sum_{\substack{XYZ}} C_{n}^{XYZ} (\mu)  \mathcal{\overline{A}}_{n}^{XYZ} (\mu)\right|^2
    \, ,\label{eq:mesondecaywidthoneloop}
\end{align}
where the spin-averaged amplitudes $\mathcal{\overline{A}}_{n}^{XYZ} (\mu)$ are the matrix elements of the renormalized operators at the
low-energy scale $\mu$: 
\begin{equation}
\mathcal{\overline{A}}_{n}^{XYZ} (\mu) = 
\langle \pi^-| \hat {\cal O}_n(\mu) | K^+ \rangle \otimes J^{XYZ}_{\ell_1 \, \ell_2} \, .
\end{equation}
Once the hadronic matrix element of the renormalized operators
is computed at the scale $\mu$ in the same regularization 
and renormalization scheme as the RGE-evolved Wilson coefficients, 
the effective lagrangian can be shown to be scale-independent, 
\begin{equation}
{\cal L}_{\rm eff} = 
\frac{1}{\Lambda_{\rm EW}^5} \, 
\sum_n C_n (\Lambda) \, 
\langle \pi^- \ell_1^+ \ell_2^+| {\cal O}_n (\Lambda) | K^+ \rangle
= 
\frac{1}{\Lambda_{\rm EW}^5} \, 
\sum_n C_n (\mu) \, 
\langle \pi^- \ell_1^+ \ell_2^+| \hat {\cal O}_n (\mu) | K^+ \rangle
\, ,
\end{equation}
where the renormalized operator is: 
\begin{equation}
\hat {\cal O}_n (\mu) = U_{nm} (\mu, \Lambda) {\cal O}_m (\Lambda) \, ,
\end{equation}
being $\hat U$ the same matrix that gives the RGE evolution of the Wilson coefficients.

For completeness, we recall that the QCD coupling constant at leading log evolves with the scale as: 
\begin{equation}
\label{eq:AlphaRun}
\alpha_s(\mu) = \frac{\alpha_s(\Lambda)}{1 - \frac{\beta_0}{2\pi} \alpha_s(\Lambda) \ln(\Lambda/\mu)},
\end{equation}
with $\beta_0 = (33 - 2 N_f)/3$ and $N_f$ the number of active quark flavors at the scale $\mu$ (either 3 or 4, depending on the mass of the decaying particle). As a starting point we took $\alpha_s(M_Z) = 0.1180(9)$ at the $Z$-boson mass~\cite{ParticleDataGroup:2024cfk}.
The numerical values of the evolution matrix $\hat U(\mu, \Lambda)$
in the $\overline{\mathrm{MS}}$ for $\mu = 2$ GeV and $\Lambda = M_Z$
are given in App.~\ref{app:umatrices}.

\section{Hadronic matrix elements at large $N_c$}
\label{sec:disconnected}

In order to explain clearly how we may estimate the hadronic matrix element under study at the next-to-leading order in large $N_c$, 
we must first introduce with the same degree of clarity how the leading contribution is computed. This will be done by shortly reviewing some basic rules of $N_c$ power-counting
and, then, introducing the {\it Vacuum Insertion Approximation}.

\subsection{Planar diagrams with soft gluon emissions at large $N_c$}
\label{sec:Ncloops}

\begin{figure}
	\centering
    \begin{tabular}{cc}
	\includegraphics[scale=0.50]{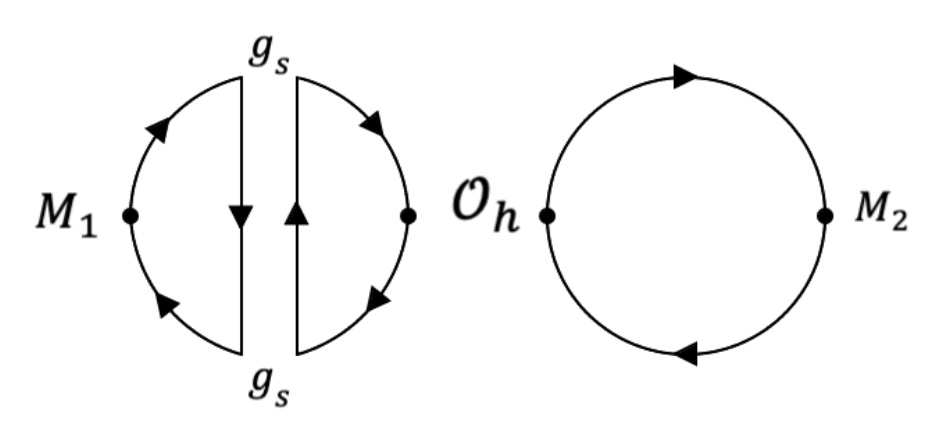} &
	\includegraphics[scale=0.50]{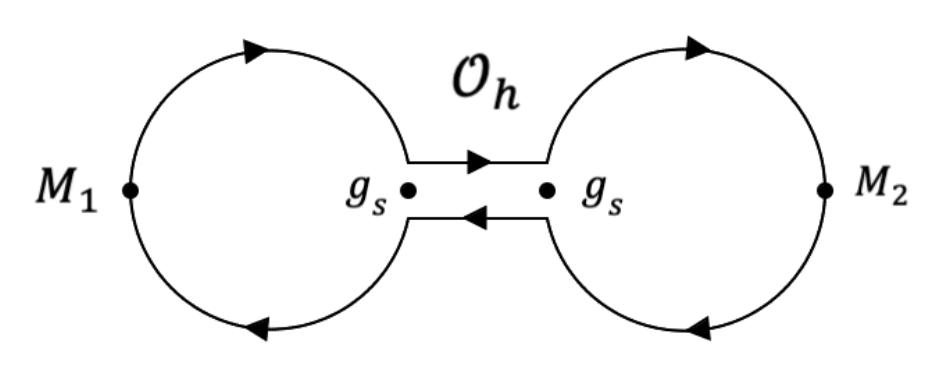} \\
	\includegraphics[scale=0.50]{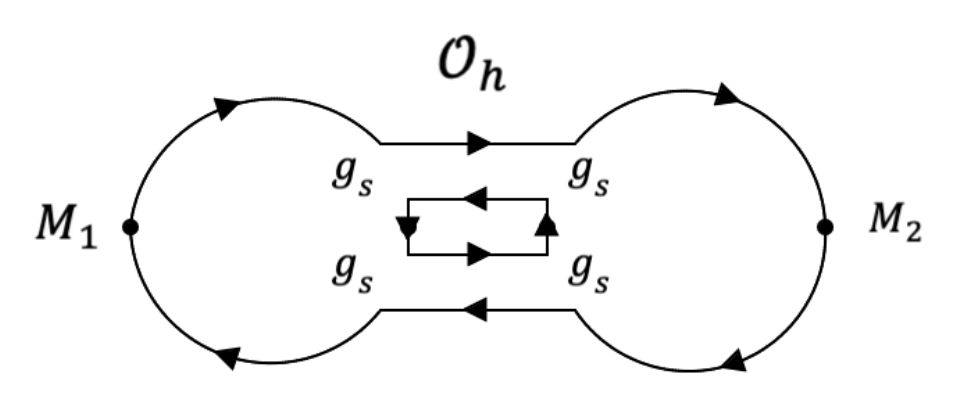} &
	\includegraphics[scale=0.50]{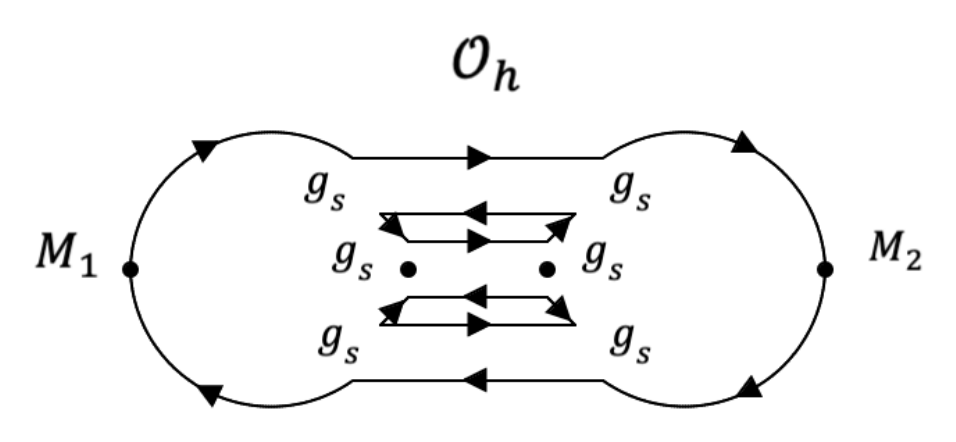}	
		\end{tabular}
    \caption{\it Disconnected diagrams with one gluon emission between initial quark lines (top left) or one (top right), two (bottom left) or three (bottom right) gluons emission
    between initial and final quark lines  for the decay $M^+_1 \to M^-_2 \ell^+ \ell^+$, with effective operator ${\cal O} = {\cal O}_h \times J_{\ell \ell}$. The top left diagram 
    is ${\cal O}(N_c^2)$, as it contains three loops and two couplings, $g_s^2 \propto 1/N_c$. On the other hand, the other three diagrams are ${\cal O}(1)$, as they contains 
    one, two and three loops and $g_s^2, g_s^4, g_s^6$ couplings, respectively. 
    }
    \label{fig:dressingdisconnected}
\end{figure}

It is well known that a quark loop (of flavour $q_i$) contributes at ${\cal O}(N_c)$ to a QCD amplitude (since quarks belong to the fundamental representation of the gauge group), 
whereas gluon loops (being in the adjoint representation) count as ${\cal O}(N_c^2)$. However,  the coupling constant $g_s$ and the number 
of colors satisfy the 't Hooft scaling in that $g_s^2 N_c = \lambda$, with $\lambda$ the 't Hooft coupling \cite{tHooft:1973alw}. This means that dressing the diagrams
in Fig.~\ref{fig:treelevel} with internal loops introduces an $N_c$-dependence that  counts both the number and type of particles in a loop and the number of QCD vertices. 
In order to understand better this point,  in Fig.~\ref{fig:dressingdisconnected} we show diagrams in which gluons are emitted between initial quark lines or
one initial and one final quark line. Gluon lines are represented as two parallel quark lines with opposite fermion number, as it is standard in the large $N_c$ expansion.
The number of QCD vertices is also depicted. In Fig.~\ref{fig:dressingdisconnected} (top left), we see a gluon emitted (and absorbed) by two initial quark lines. 
Since a gluon is represented by two (anti)parallel quark lines, we see that the initial loop splits into two loops and the final loop is untouched. As a result of the gluon emission, three
loops are present, that therefore count as ${\cal O} (N_c^3)$. However, two QCD couplings are also present, and thus $g_s^2 \propto 1/N_c$ at fixed 't Hooft coupling $\lambda$. 
Eventually, emission of a gluon between initial quark lines give a diagram that is ${\cal O}(N_c^2)$, the same as the original disconnected tree-level diagram. The same would happen if the gluon is emitted between two final quarks lines. On the other hand, we show in Fig.~\ref{fig:dressingdisconnected} (top right) the emission of a gluon between one initial and one final quark line. In this case, only one closed loop is present, as the two disconnected ones become color-connected. Therefore, the power-counting gives $N_c$ (one closed loop) times $1/N_c$ (due to two couplings $g_s^2$), such that this diagram is ${\cal O}(1)$. It can be seen that, at large $N_c$, the second diagram is ${\cal O}(1/N_c^2)$ with respect to the first one (and to the tree-level one). The same result is obtained for  Fig.~\ref{fig:dressingdisconnected} (bottom left) and  Fig.~\ref{fig:dressingdisconnected} (bottom right), where
the emission of two and three gluons between initial and final quark lines is shown. Also in these two cases, the number of color loops is the same as the number of $g_s^2$ insertions and, 
thus, all of these diagrams are ${\cal O}(1)$.  

\begin{figure}
	\centering
    \begin{tabular}{cc}
	\includegraphics[scale=0.50]{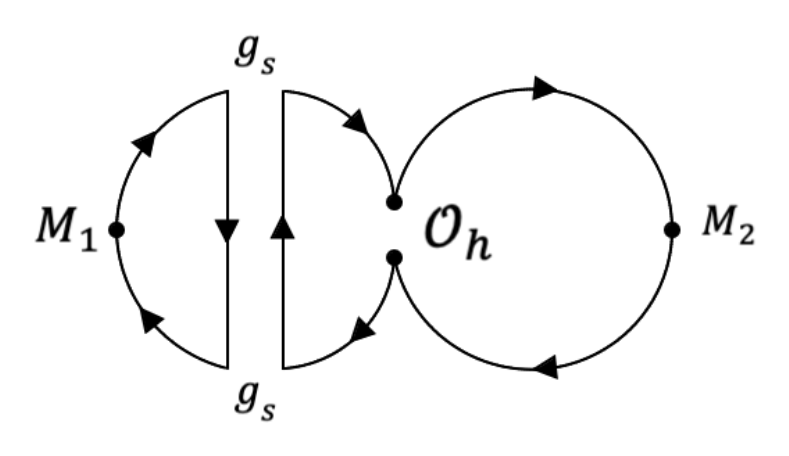} &
	\includegraphics[scale=0.55]{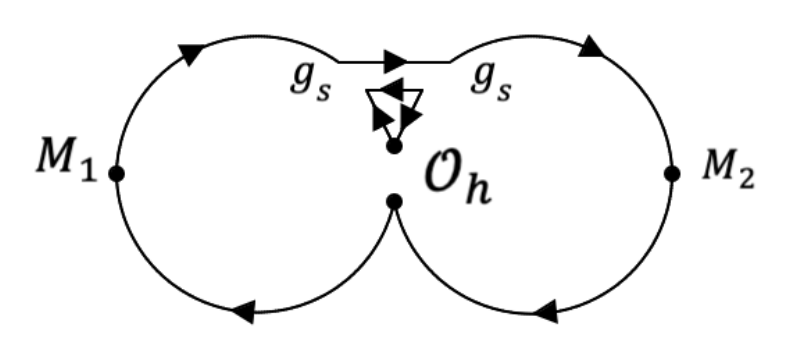}
		\end{tabular}
    \caption{\it Connected diagrams with one gluon emission between initial quark lines (left) or 
    between initial and final quark lines (right)  for the decay $M^+_1 \to M^-_2 \ell^+ \ell^+$, with effective operator ${\cal O} = {\cal O}_h \times J_{\ell \ell}$. Both diagrams 
    are ${\cal O}(N_c)$, as they contain two loops and two couplings, $g_s^2 \propto 1/N_c$.}
    \label{fig:dressingconnected}
\end{figure}

Consider now the tree-level diagram in Fig.~\ref{fig:treelevel} (right). In this case, we have a single loop and no QCD couplings. Therefore, at large $N_c$ this diagram
counts as ${\cal O}(N_c)$. In Fig.~\ref{fig:dressingconnected} we show gluon emissions between initial (or final) quark lines (left) or between one initial and one final quark
line (right) in the case of the connected diagram. It is easy to see that both diagrams are ${\cal O} (N_c)$, as they have two loops and two QCD couplings, $g_s^2 \propto 1/N_c$. 
Therefore, we can classify the different diagrams as follows: 
\begin{itemize}
\item {\bf Leading}: Disconnected diagrams (tree-level and with gluons between either initial or final quarks), ${\cal O} (N_c^2)$;
\item {\bf Sub-leading}: Connected diagrams (tree-level and dressed), ${\cal O} (N_c)$;
\item {\bf Sub-sub-leading}: Disconnected diagrams with gluons between initial and final quarks, ${\cal O}(1)$. 
\end{itemize}

\subsection{The Vacuum Insertion Approximation}
\label{sec:VIA}

 Since carrying out the non-perturbative computation is usually a hard task, if we consider only the leading contribution at large $N_c$, we may get a first estimate of the hadronic matrix element. At leading order in $N_c$, for the $K^+ \to \pi^- \, \ell^+ \ell^+$ process we have:
\begin{equation}
\langle \pi (z) | \bar u (x) \Gamma d (x) \bar u (x) \Gamma^\prime s(x) | K (0) \rangle = 
                         \sum_n \langle \pi (z) | \bar u (x) \Gamma d (x) | n \rangle \langle n | \bar u (x) \Gamma^\prime s(x) | K (0) \rangle + {\cal O} (1/N_c) \, ,
\end{equation}
where $\Gamma, \Gamma^\prime$ are some of the $\gamma$-matrices in the operator basis of eq.~(\ref{eq:relevantoperators}). As the leading contribution at large $N_c$
is represented by disconnected diagrams, we have inserted a complete set of intermediate states $|n\rangle$ and summed over them. The set of intermediate states takes into
account that, non-perturbatively, any possible state compatible with the quantum numbers of $J_h | K^+ \rangle$ and $J^\prime_h | \pi^- \rangle$ should be produced in principle. 
Of all the possible states that can be produced, the lightest is of course $|0\rangle$. On the other hand, if we produce a meson state $| M \rangle \langle M |$, we should
recall that at large $N_c$ any source of a meson field counts as $1/\sqrt{N_c}$ \cite{Witten:1978bc,Witten:1979kh}. This means that producing and annihilating an intermediate meson state $|n\rangle$,
different from the vacuum, corresponds to a sub-leading contribution at large $N_c$. 

As a consequence, at large $N_c$ the leading contribution to the hadronic matrix element comes from the disconnected diagram in which we only consider the vacuum 
as a possible intermediate state. The so-called {\it Vacuum Insertion Approximation} (VIA) states, then, that: 
\begin{eqnarray}
\langle \pi (z) | \bar u (x) \Gamma d (x) \bar u (x) \Gamma^\prime s(x) | K (0) \rangle = 
                        \langle \pi (z) | \bar u (x) \Gamma d (x) | 0 \rangle \langle 0 | \bar u (x) \Gamma^\prime s(x) | K (0) \rangle + {\cal O} (1/N_c) \, .
                       \nonumber \\
\end{eqnarray}
The two resulting two-point correlation functions, $\langle \pi^- | \bar u (x) \Gamma d (x) | 0 \rangle$ and $\langle 0 | \bar u (x) \Gamma^\prime s(x) | K (0) \rangle$,
can be estimated non-perturbatively using soft-pion theorems, as is reminded in Ref.~\cite{Gonzalez:2023him}: 
 \begin{equation}
 \label{eq:PCACrelations}
 \left \{
\begin{array}{l}
\langle 0 | \bar q_1 \gamma_5 q_2 | M \rangle = i \xi_M f_M \, , \\
\\
\langle 0 | \bar q_1 \gamma_\mu \gamma_5 q_2 | M \rangle = i p_\mu  f_M \, ,
\end{array}
 \right .
 \end{equation}
 with $\xi_M = m_M^2/(m_1 + m_2)$ the pseudo-scalar density of the meson $M$ (with $m_1, m_2$ the masses of the corresponding constituent quarks), $f_M$ the meson decay constant and $p_\mu$ the meson momentum. On the other hand, being the vacuum a singlet under the Lorentz group, the matrix element of the tensor current vanishes:
\begin{equation}
\label{eq:PCACtensor}
\langle 0 | \bar q_1\sigma^{\mu\nu} P_X q_2 | M \rangle = 0 
\end{equation}
due to Lorentz invariance (see also Ref.~\cite{deGouvea:2007qla}). 

In our computations, we have  used $f_{K^\pm} = 155.7 \pm 0.7$ MeV and $f_{\pi^\pm} = 130.2 \pm 0.8$ MeV (from the most recent lattice results, see Ref.~\cite{FlavourLatticeAveragingGroupFLAG:2024oxs})
and the PDG values of the quarks running masses in the  $\overline{\mathrm{MS}}$ scheme at $\mu = 2$ GeV (see Ref.~\cite{ParticleDataGroup:2022pth}),
$m^{\overline{\mathrm{MS}}}_u(2 \; {\rm GeV}) = 2.16^{+0.49}_{-0.26}$ MeV,  
$m^{\overline{\mathrm{MS}}}_d(2 \; {\rm GeV}) = 4.67_{-0.17}^{+0.48}$ MeV and 
$m^{\overline{\mathrm{MS}}}_s(2 \; {\rm GeV}) = 93.4_{-3.4}^{+8.6}$ MeV.
 For the operator ${\cal O}_h = {\cal O}_3^{XXZ}$ 
 in the type-I see-saw model, this is how we obtained the expression in eq.~(\ref{eq:schannelseesaw}). Notice that applying soft-pion theorems to the two-point correlation functions above gives a non-perturbative computation of each disconnected loop, which is encoded in a non-perturbative parameter to be measured experimentally, as in the case of $f_\pi$ and $f_K$. In this approximation, the only scale dependence of the hadronic matrix elements comes through the quark running masses (as, due to the PCAC, the decay constants do not run). 
 For the general case at hand, we may derive an estimate
 for the amplitudes $\mathcal A^{XYZ}_i$ valid at leading order
 in $1/N_c$:
\begin{align}
    \mathcal  A_{1d}^{XYZ} & = \pm \, 2 \, f_{M_1} \, f_{M_2} \, 
    \xi_{M_1} \, \xi_{M_2} 
    \otimes 
    [\bar u(p_{\ell_1}) P^Z\, v(p_{\ell_2})] 
    + {\cal O}(1/N_c)
    \, ,\label{eq:A1}\\
    \nonumber \\
    \mathcal  A_{3d}^{XYZ} & = \pm \, 2 \, f_{M_1} \, f_{M_2} \, 
    (p_M\cdot p_M') \, 
    \otimes 
    [\bar u(p_{\ell_1}) P^Z\, v(p_{\ell_2})] 
    + {\cal O}(1/N_c) \, ,\\
    \nonumber\\
    \mathcal A_{5d}^{XYZ}, \mathcal A_{5d}^{\prime \, XYZ}  
    & = \pm \, 2 \, f_{M_1} \, f_{M_2} \, 
    \xi_{M_1} \, p_M^\mu 
    \otimes 
    [\bar u(p_{\ell_1}) \gamma_\mu P^Z\, v(p_{\ell_2})]
    + {\cal O}(1/N_c) \, ,\\
    \nonumber\\
    \mathcal A_{6d}^{XYZ} & = \pm \, 2 \, f_{M_1} \, f_{M_2} \, 
    p_{M_1}^\mu \, p_{M_2}^\nu \otimes 
    [\bar u(p_{\ell_1}) \sigma_{\mu\nu} P^Z \, v(p_{\ell_2})] 
    + {\cal O}(1/N_c) \, , \label{eq:A6}
\end{align}
where $+$ and $-$ correspond to cases $X=Y$ and $X\neq Y$, respectively, and the label $d$ reminds us that the amplitudes only
take into account the contribution of the disconnected diagram. 
Due to eq.~(\ref{eq:PCACtensor}), 
the amplitudes ${\cal A}_{2d},{\cal A}_{4d},{\cal A}_{7d}$ and ${\cal A}_{8d}$ vanish in this approximation for any choice of the chiralities $X,Y, Z$, both for identical or different final leptons.

\subsection{Bounds on $C_n(\mu)$ at leading order in $1/N_c$}

Once we have an estimate of the hadronic matrix elements at leading order in $1/N_c$, we may use eq.\,(\ref{eq:mesondecaywidthtreelevel}) to find tree-level bounds on the Wilson coefficients $C_{n (q_1 q_2) (q_3 q_4)}^{XYZ}$. 
The bounds on the tree-level Wilson coefficients are shown in Sect.~\ref{sec:results}, in the second column of Tabs.~\ref{tab:results1} and 
\ref{tab:results2} (for $\ell_1 = \ell_2 = e$ and $\mu$, respectively) and \ref{tab:results3} (for $\ell_1 = e$ and $\ell_2 = \mu$). 
Notice that no tree-level bound can be put on the Wilson coefficients $C_2, C_4, C_7$ and $C_8$, as the corresponding amplitudes, $\mathcal A_{2d}$, $\mathcal A_{4d}$, $\mathcal A_{7d}$, and $\mathcal A_{8d}$, vanish at the leading order in $1/N_c$, since meson decay modes mediated by tensor currents are suppressed \cite{deGouvea:2007qla,Quintero:2016iwi}.

Dressing the relevant operators with hard gluons and including the
corresponding one-loop QCD corrections in our amplitudes, we can use the RG evolution matrices $\hat U(\mu,\Lambda)$, computed from 
eq.~(\ref{eq:UMatrix}), to derive RGE-improved limits on the Wilson coefficients. 

If, using eq.~(\ref{eq:mesondecaywidthtreelevel}), we were able
to put an experimental bound on the tree-level Wilson coefficient 
$C_n \lesssim C_n^{exp}$ then, due to radiative corrections, 
we must compute the RGE-evolved Wilson coefficient $C_n(\mu)$
and apply the constraint as follows:
\begin{align}
C_n \to C_n(\mu) = \sum_{m}\hat U_{nm}(\mu,\Lambda) C_m(\Lambda) \lesssim C_n^{exp},\label{eq:newbounds}
\end{align}
where the indices $n,m$ run over the operators that are mixed by the evolution matrix $\hat U(\mu,\Lambda)$. In the absence of a non-perturbative computation of the hadronic matrix elements, we may again make use of the VIA at the leading order in $1/N_c$ to get an estimate
of the experimental bounds on the RGE-evolved Wilson coefficient.
This is done using eq.~(\ref{eq:mesondecaywidthoneloop}). 
We give the RGE-improved bounds for the Wilson coefficients in Sect.~\ref{sec:results},
in the third column of Tab.~\ref{tab:results1}, \ref{tab:results2} (for identical final leptons) and \ref{tab:results3} (for different final leptons). 

The main impact of the RGE-running of the Wilson coefficients from the high-energy scale $\Lambda$ to the low-energy scale $\mu$ is that
we are able to put constraints on the Wilson coefficients $C_2, C_4, C_4^\prime, C_6, C_7, C_7^\prime$, something impossible
using the tree-level expressions. This is because, through the evolution matrix $\hat U$, the matrix element of the renormalized operator $\hat {\cal O}(\mu)$ is: 
\begin{equation}
\label{eq:renormalizedamplitudes}
\left [{\cal A}_n^{XYZ} (\mu) \right ]_d = 
\sum_m U_{nm} (\mu, \Lambda) \,  
\left [ {\cal A}_m^{XYZ} (\Lambda) \right ]_d \, ,
\end{equation}
where the right-hand side can be computed in the VIA using 
eqs.~(\ref{eq:A1})-(\ref{eq:A6}).

\section{Connected diagrams contribution to $K^+ \to \pi^- l^+ l^+$ $\Delta L = 2$ transitions}
\label{sec:connected}

After computing the perturbative one-loop renormalization of the operators in the basis shown in  eq.~(\ref{eq:relevantoperators}) (Sect.~\ref{sec:perturbative}) and giving a 
non-perturbative estimate of the hadronic matrix elements of those operators applying the VIA at the leading order in $1/N_c$, 
 we may now compute the long-range sub-leading contribution at large $N_c$ to the hadronic matrix elements. Recall that the amplitude corresponding to inserting a 
 quark four-fermion operator ${\cal O}_h$ between the initial and final meson states $| K^+ \rangle$ and 
$| \pi^- \rangle$ can be obtained by the following ${\cal T}$-product: 
\begin{eqnarray}
\label{eq:Tproduct}
\langle \pi^- (z) | {\cal O}^{XYZ}_i (x) | K^+ (0) \rangle &=& {\cal T} \left [
\langle 0 | \left ( \bar d \gamma_5 u \right )_z \, \left ( \bar u \Gamma_X d \right )_x \left ( \bar u \Gamma_Y s \right )_x
\left ( \bar s \gamma_5 u \right )_0 | 0 \rangle   \right ] \\
&& \nonumber \\
& = & 
\langle 0 |
  \tikzmarknode{N1}{\bar d} (z) \gamma_5
    \tikzmarknode{N2}{u} (z) \, 
    \tikzmarknode{N3}{\bar u} (x) \Gamma_X
    \tikzmarknode{N4}{d} (x) \; 
      \tikzmarknode{N5}{\bar u} (z) \Gamma_Y
    \tikzmarknode{N6}{s} (z) \, 
    \tikzmarknode{N7}{\bar s} (0) \gamma_5
    \tikzmarknode{N8}{u} (0) | 0 \rangle 
    \begin{tikzpicture}[overlay,remember picture,shorten <=2pt]
        \draw (N1) |-++ (3.2,.5) node[right] (V) {};
        \draw (N4) |- (V);
        \draw (N2) |-++ (.8,-.5) node[right] (R) {};
        \draw (N3) |- (R);
        \draw (N5) |-++ (3.2,.5) node[right] (V) {};
        \draw (N8) |- (V);
        \draw (N6) |-++ (.8,-.5) node[right] (R) {};
        \draw (N7) |- (R);
    \end{tikzpicture}
    \nonumber \\
    && \nonumber \\
    &-&
\langle 0 |
  \tikzmarknode{N1}{\bar d} (z) \gamma_5
    \tikzmarknode{N2}{u} (z) \, 
    \tikzmarknode{N3}{\bar u} (x) \Gamma_X
    \tikzmarknode{N4}{d} (x) \; 
      \tikzmarknode{N5}{\bar u} (z) \Gamma_Y
    \tikzmarknode{N6}{s} (z) \, 
    \tikzmarknode{N7}{\bar s} (0) \gamma_5
    \tikzmarknode{N8}{u} (0) | 0 \rangle \, .
    \begin{tikzpicture}[overlay,remember picture,shorten <=2pt]
        \draw (N1) |-++ (3.2,.5) node[right] (V) {};
        \draw (N4) |- (V);
        \draw (N2) |-++ (3,-.5) node[right] (R) {};
        \draw (N5) |- (R);
        \draw (N3) |-++ (5.4,-.7) node[right] (V) {};
        \draw (N8) |- (V);
        \draw (N6) |-++ (.8,.5) node[right] (R) {};
        \draw (N7) |- (R);
    \end{tikzpicture}
    \nonumber \\
    && \nonumber
    \end{eqnarray}
The first amplitude corresponds to the disconnected diagram, and it can be directly approximated inserting a complete set of intermediate states $|n\rangle \langle n|$ and then applying the VIA (see Sect.~\ref{sec:disconnected}). On the other hand, the second amplitude corresponds to the connected diagram and the VIA cannot be applied directly. The relative sign between the two comes from the different number of closed fermion loops. 

In order to get a VIA-consistent estimate of the hadronic matrix element up to subleading order in $1/N_c$, we must relate the connected diagram 
(unaccounted for up to now) to the disconnected diagram, whose computation in the VIA gives the $N_c \to \infty$ limit and was studied in the previous Section. This is usually done by means of Fierz-rearranging the fields in the local effective operator, so as to relate one kind of contractions with the external states (that give rise to the connected diagram) to the other one (that give rise to the disconnected diagram). By doing this, operators belonging to the
basis in eq.~(\ref{eq:relevantoperators}) may mix among themselves. Quite generally, $1/N_c$ factors arise due to the corresponding rearrangement of color indices. This is a straightforward consequence of the fact that the connected diagram is ${\cal O}(N_c)$ in the large $N_c$ expansion, whereas the disconnected diagram is ${\cal O}(N_c^2)$. 

The Fierz-rearranging rules of spinor indices in Minkowski space-time for the four-fermion operators relevant for the $K^+ \to \pi^- \ell^+ \ell^+$ transition at hand are given in App.~\ref{app:fierz}.  Notice that the anti-commutation of fermion fields in the ${\cal T}$-product is already considered.
After performing spin-indices rearrangement, a color-indices rearrangement must be applied, too, using the relation: 
 \begin{equation}
 \label{eq:colorfierz}
\delta_{AB} \otimes \delta_{CD}  = - \frac{1}{2 N_c} \, \delta_{AD} \otimes \delta_{CB}  + \frac{1}{2} \, \sum_a t^a_{AD} \otimes t^a_{CB},
 \end{equation}
with $a = 1, \dots, N_c^2 - 1$ color indices in the adjoint representation and $A, B, C, D = 1, \dots, N_c$ color indices in the fundamental representation, respectively.
Eventually, after rearranging spinor and color indices, we can give an estimate of the non-perturbative amplitudes in terms of the ${\cal O}(1/N_c)$ mixing matrix $\hat B$: 
\begin{equation}
\label{eq:Bmatrix}
\left [\vec {\cal A}^{XYZ} \right ]_{\rm VIA} = \hat B \left ( \frac{1}{N_c} \right ) \, 
\left [\vec {\cal A}^{XYZ} \right ]_d \, 
\end{equation}
being the disconnected amplitudes $\vec {\cal A}^{XYZ}_d$
given in eqs.~(\ref{eq:A1})-(\ref{eq:A6}), 
for the two cases of identical chiralities in the hadronic
currents ($XXZ$), or different chiralities ($XYZ$, with $X \neq Y$). 

The computation of the sum of connected and disconnected diagrams
in the VIA is given in App.~\ref{app:sums}. Using, then, the relations in eqs.~(\ref{eq:PCACrelations}) and (\ref{eq:PCACtensor}), it is possible to reduce the non-perturbative amplitude $\langle M_2 | {\cal O}_i^{XYZ} | M_1 \rangle $ to simple expressions that depend on the meson decay constants and on the pseudoscalar densities, only (such as those in eq.~(13-16) of Ref.~\cite{Gonzalez:2023him}), albeit including ${\cal O}(1/N_c)$
corrections induced by the connected diagrams. 
As we have seen above, the connected diagrams may induce mixing
between different operators belonging to the basis in eqs.~(\ref{eq:relevantoperators}) and (\ref{eq:relevantoperators2}). 
In the case of identical chiralities for the two hadronic currents, we have: 
\begin{equation}
\label{eq:VIAllXX}
\left (
\begin{array}{c}
{\cal A}_1^{XXZ} \\
\\
{\cal A}_2^{XXZ} \\
\\
{\cal A}_3^{XXZ} \\
\\
{\cal A}_4^{XXZ} \\
\\
{\cal A}_4^{\prime \, XXZ} \\
\\
{\cal A}_5^{XXZ} \\
\\
{\cal A}_5^{ \prime \, XXZ}
\end{array}
\right )_{\rm VIA} = 
\left (
\begin{array}{ccccccc}
1 - \frac{1}{4 N_c} & 0 & 0 & 0 & 0 & 0 & 0 \\
\\
\left ( - \frac{3}{N_c} \right )^\star & 0 & 0 & 0 & 0 & 0 & 0 \\
\\
0 & 0 & 1 + \frac{1}{2 N_c} & 0 & 0 & 0 & 0 \\
\\
0 & 0 & 0 & 0 & 0 & 0 & - \frac{3 \, i}{4 N_c} \\
\\
0 & 0 & 0 & 0 & 0 &  - \frac{3 \, i}{4 N_c} & 0 \\
\\
0 & 0 & 0 & 0 & 0 & 1 & - \frac{1}{4 N_c} \\
\\
0 & 0 & 0 & 0 & 0 & - \frac{1}{4 N_c} & 1 \\
\end{array}
\right ) \, 
\left (
\begin{array}{c}
{\cal A}_1^{XXZ} \\
\\
{\cal A}_2^{XXZ} \\
\\
{\cal A}_3^{XXZ} \\
\\
{\cal A}_4^{XXZ} \\
\\
{\cal A}_4^{\prime \, XXZ} \\
\\
{\cal A}_5^{XXZ} \\
\\
{\cal A}_5^{ \prime \, XXZ}
\end{array}
\right )_d \, , 
\end{equation}
where the $^\star$ near the matrix element $B_{21}$, that
gives the relation between ${\cal A}_2^{XXZ}$ in the VIA 
with ${\cal A}_{1d}^{XXZ}$, indicates that it  
only differs from zero for $Z = X$.
In the case of operators with different chiralities, we have: 
\begin{equation}
\label{eq:VIAllXY}
\left (
\begin{array}{c}
{\cal A}_1^{XYZ} \\
\\
{\cal A}_2^{XYZ} \\
\\
{\cal A}_3^{XYZ} \\
\\
{\cal A}_4^{XYZ} \\
\\
{\cal A}_4^{\prime \, XYZ} \\
\\
{\cal A}_5^{XYZ} \\
\\
{\cal A}_5^{ \prime \, XYZ}
\end{array}
\right )_{\rm VIA} = 
\left (
\begin{array}{ccccccc}
1  & 0 & - \frac{1}{4 N_c} & 0 & 0 & 0 & 0 \\
\\
0 & 0 & 0 & 0 & 0 & 0 & 0 \\
\\
- \frac{1}{ N_c} & 0 & 1 & 0 & 0 & 0 & 0 \\
\\
0 & 0 & 0 & 0 & 0 & \frac{3 \, i}{4 N_c} & 0 \\
\\
0 & 0 & 0 & 0 & 0 & 0 & \frac{3 \, i}{4 N_c}  \\
\\
0 & 0 & 0 & 0 & 0 & 1 - \frac{1}{4 N_c} & 0 \\
\\
0 & 0 & 0 & 0 & 0 & 0 & 1 - \frac{1}{4 N_c}  \\
\end{array}
\right ) \, 
\left (
\begin{array}{c}
{\cal A}_1^{XYZ} \\
\\
{\cal A}_2^{XYZ} \\
\\
{\cal A}_3^{XYZ} \\
\\
{\cal A}_4^{XYZ} \\
\\
{\cal A}_4^{\prime \, XYZ} \\
\\
{\cal A}_5^{XYZ} \\
\\
{\cal A}_5^{ \prime \, XYZ}
\end{array}
\right )_d \, . 
\end{equation}
The amplitudes of the operators ${\cal O}_6^{XYZ}, {\cal O}_7^{XYZ}, 
{\cal O}_7^{\prime \, XYZ}$ 
and ${\cal A}_8^{XYZ}$ mostly do not mix under Fierzing. We have: 
\begin{equation}
\label{eq:VIAl1l26}
\left \{
\begin{array}{l}
\left [ {\cal A}_6^{XXX}\right ]_{\rm VIA} = 
 \left ( 1 + \frac{1}{2 N_c}\right ) \, {\cal A}_{6d}^{XXX} \, ,\\
\\
\left [ {\cal A}_6^{XXZ}\right ]_{\rm VIA} = 
\left ( 1 - \frac{1}{2 N_c}\right ) \, {\cal A}_{6d}^{XXZ} \, ,\\
\\
\left [ {\cal A}_6^{XYX}\right ]_{\rm VIA} = 
{\cal A}_{6d}^{XYX} \, ,
\end{array}
\right .
\end{equation}
and 
\begin{equation}
\label{eq:VIAl1l278}
\left \{
\begin{array}{l}
\left [ {\cal A}_7^{XXX}\right ]_{\rm VIA} = 0 \, ,\\
\\
\left [ {\cal A}_7^{\prime \, XXX}\right ]_{\rm VIA} = 0 \, ,\\
\\
\left [ {\cal A}_8^{XXX}\right ]_{\rm VIA} = 0 \, .\\
\end{array}
\right . 
\end{equation}
The only exception is the case of operators with $XYY$ chiralities, for which: 
\begin{equation}
\label{eq:VIAl1l267XY}
\left (
\begin{array}{c}
{\cal A}_6^{XYY} \\
\\
{\cal A}_7^{XYY} \\
\\
{\cal A}_7^{\prime \, XYY} \\
\end{array}
\right )_{\rm VIA} = 
\left (
\begin{array}{ccc}
1 & 0 & 0 \\
\\
-\frac{i}{N_c} & 0 & 0 \\
\\
\frac{i}{N_c} & 0 & 0
\end{array}
\right ) \, 
\left (
\begin{array}{c}
{\cal A}_6^{XYY} \\
\\
{\cal A}_7^{XYY} \\
\\
{\cal A}_7^{\prime \, XYY} \\
\end{array}
\right )_{d}
\end{equation}

The main result in eqs.~(\ref{eq:VIAllXX})-(\ref{eq:VIAl1l267XY}) is
that we may improve the bounds on the Wilson coefficients
$C_2(\mu), C_4(\mu), C_7(\mu)$ and $C_8(\mu)$ using the ${\cal O}(1/N_c)$-improved VIA, due to the mixing of the corresponding operators
with those for which the amplitudes ${\cal A}_d^{XYZ}$ are non-vanishing.

\section{Improved bounds on the Wilson coefficients}
\label{sec:results}

In this section, we eventually present our ${\cal O}(1/N_c)$ improved bounds on the Wilson coefficients corresponding to the operator basis defined in eqs. (\ref{eq:relevantoperators}-\ref{eq:relevantoperators2}), by taking into account the contributions from the connected diagrams in the hadronic matrix elements in 
the {\it Vacuum Insertion Approximation}. 
In order to derive the bounds, we 
consider the effect of one operator at a time.
The bounds are obtained from the most recent experimental limits \cite{NA62:2022tte,NA62:2019eax,NA62:2021zxl}:
\begin{align}
    \text{Br}(K^+ \to \pi^- e^+ e^+) \leq 5.3\times 10^{-11},\nonumber\\
    \text{Br}(K^+ \to \pi^- \mu^+ \mu^+) \leq 4.2\times 10^{-11},\\
    \text{Br}(K^+ \to \pi^- e^+ \mu^+) \leq 4.2\times 10^{-11},\nonumber
\end{align}
and are shown in Tabs.~(\ref{tab:results1}) and (\ref{tab:results2}),
for identical final state leptons ($e^+ e^+$ and $\mu^+ \mu^+$, respectively)
and (\ref{tab:results3}) for different final state leptons.
In each Table, the first column lists the considered Wilson coefficient, 
numbered as their corresponding operator, with an explicit reference
to the chiralities of the two quark currents (in the case of different final leptons, Tab.~\ref{tab:results3}, when needed also the chirality of the lepton current is given); the second column presents the tree-level bounds obtained from the expression for the decay width, eq.~(\ref{eq:mesondecaywidthtreelevel}); the third column includes the effect of (perturbative) operator mixing under RGE evolution, as developed in Ref.~\cite{Gonzalez:2023him} and reviewed in Sects.~\ref{sec:perturbative} and \ref{sec:disconnected}; and, finally, 
the fourth column shows the bounds incorporating both RGE evolution effects and the (non-perturbative) contribution of connected diagrams to the hadronic matrix elements, presented in Sect.~\ref{sec:connected}.

\begin{table}[h]
\centering
\begin{tabular}{|c|c|c|c|}
\hline
&&& \\
\textbf{Coefficient} & \textbf{Tree} & \textbf{RGE-Imp} \cite{Gonzalez:2023him} & \textbf{RGE + connected diag.} \\
&&& \\
\hline
\hline
$C_{1}^{XXZ}$     & $2.5\times 10^2$ & $1.5\times 10^2$ & $1.2\times 10^2$\\
\hline
$C_{1}^{XYZ}$     & $2.5\times 10^2$ & $1.1\times 10^2$ & $1.1\times 10^2$ \\
\hline
$C_{2}^{XXX}$     & --            & $2.8\times 10^4$ & $4.5\times 10^2$ \\
\hline
$C_{3}^{XXZ}$     & $1.9\times 10^4$ & $2.3\times 10^4$ & $2.0\times 10^4$ \\
\hline
$C_{3}^{XYZ}$     & $1.9\times 10^4$ & $2.1\times 10^4$ & $8.5\times 10^2$             \\
\hline
$C_{4}^{XXZ}$     & --            & $3.0\times 10^4$ & $9.2\times 10^3$ \\
\hline
$C_{4}^{\prime XXZ}$    & --            & $3.3\times 10^4$ & $1.1\times 10^4$ \\
\hline
$C_{4}^{XYZ}$     & --            & $1.3\times 10^5$ & $1.3\times 10^4$ \\
\hline
$C_{4}^{\prime XYZ}$    & --            & $1.2\times 10^5$ & $1.1\times 10^4$ \\
\hline
$C_{5}^{XXZ}$     & $2.3\times 10^3$ & $1.5\times 10^3$ & $1.7\times 10^3$             \\
\hline
$C_{5}^{\prime XXZ}$    & $2.1\times 10^3$ & $1.4\times 10^3$ & $1.5\times 10^3$             \\
\hline
$C_{5}^{XYZ}$     & $2.3\times 10^3$ & $1.8\times 10^3$ & $1.9\times 10^3$ \\
\hline
$C_{5}^{\prime XYZ}$    & $2.1\times 10^3$ & $1.6\times 10^3$ & $1.7\times 10^3$ \\
\hline
\hline
\end{tabular}
\caption{\it Bounds on Wilson coefficients at $\mu = 2$ GeV 
from the process $K^+\to \pi^-e^+e^+$: at tree level, including RGE, and with the effect of connected diagrams. The labels $X, Y$ stand for the chiralities of the two quark bilinears, whereas $Z$ refers to
the chirality of the lepton one, respectively. When the two quark-related labels are identical ($XX$), the bound applies to both $LL,RR$ operators. When they differ ($XY$), the bound applies to $LR, RL$. On the other hand, the chirality of the lepton bilinear is irrelevant but for $C_2^{XXX}$, for which the bound applies only for the combinations
$LLL, RRR$.
}
\label{tab:results1}
\end{table}

We will use Tab.~\ref{tab:results1} to explain in detail our results. 
First of all notice that, at tree-level, only some Wilson coefficients can be bounded. This is because some of the hadronic matrix elements vanish, when approximated by their disconnected diagram contribution neglecting soft gluon interchange between the initial and final meson ({\it i.e.}, in the VIA at leading order in $1/N_c$). This is indeed the case of the operators
${\cal O}_2, {\cal O}_4$ and ${\cal O}_4^\prime$ (for identical final leptons), together with ${\cal O}_7, {\cal O}_7^\prime$ and ${\cal O}_8$ (for different final leptons). 
The tree-level bounds at the high-energy scale $\Lambda$ (that in this work is taken to be $M_Z$) are rather loose: $O_1$ has a Wilson coefficient $C_1(M_Z) \leq 2.5 \times 10^2$; ${\cal O}_5$ has a Wilson coefficient $C_5(M_Z) \leq 2.3 \times 10^3$; ${\cal O}_5^\prime$ has a Wilson coefficient $C_5^\prime (M_Z) \leq 2.1 \times 10^3$; and, eventually, ${\cal O}_3$ has a Wilson coefficient $C_3 (M_Z) \leq 1.9 \times 10^4$. In all cases, the chiralities of the two quark currents are irrelevant: the same bounds are obtained for LL, RR, LR or RL (irrespectively of the chirality of the lepton current). Notice that the operator ${\cal O}_3$ (the one we would get in a theory whose new physics only involves left-handed vector currents, as in the SM) has the loosest bound. 

After running the Wilson coefficients down to the scale where the hadronic matrix elements should be computed (for which we choose $\mu = 2$ GeV), we can see that the effect of perturbative mixing induced by
one-loop RGE is extremely significant: due to mixing, we can indeed constrain operators whose tree-level Wilson coefficients were unbounded.
This is the case of ${\cal O}_2$, for which we get $C_2 (2 \; $\rm GeV$) \leq 2.8 \times 10^4$;  ${\cal O}_4$, for which we get\footnote{For the operators ${\cal O}_4$ and ${\cal O}_4^\prime$, chirality matters: we get different bounds depending on whether we have identical (LL or RR) or different (LR or RL) chiralities for the quark currents.} $C_4 (2 \, $\rm GeV$) \leq 3.0 \times 10^4$ ($1.3 \times 10^5$); and ${\cal O}_4^\prime$, for which we get $C_4^\prime (2 \, $\rm GeV$) \leq 3.3 \times 10^4$ ($1.2 \times 10^5$). 
The other operators are also significantly modified. RGE-improved bounds on ${\cal O}_1, {\cal O}_3, {\cal O}_5$ and ${\cal O}_5^\prime$ are: $C_1(2 \, {\rm GeV}) \leq 1.5 \times 10^2$ ($1.1 \times 10^2$); $C_5(2 \, {\rm GeV}) \leq 1.5 \times 10^3$ ($1.8 \times 10^3$); and $C_5^\prime(2 \, {\rm GeV}) \leq 1.4 \times 10^3$ ($1.6 \times 10^3$)
for identical (different) quark bilinear chiralities. 
We can see that the difference 
ranges from 50\% to 300\% (in the case of $C_1^{XYZ}$). The only operator
for which the RGE from $M_Z$ down to $\mu = 2$ GeV loosen the tree-level bound is ${\cal O}_3$, for which we get $C_3(2 \, {\rm GeV}) \leq 2.3 \times 10^4$ ($2.1 \times 10^4$), depending of the chiralities of the quark currents ($XX$ or $XY$, with $X \neq Y$, respectively).

Eventually, in the last column we show the impact on the Wilson coefficients of the contribution of the connected diagrams. These diagrams, although subleading in $1/N_c$, must be added to the disconnected ones in order to reconstruct consistently the full hadronic amplitude in a controlled approximation (the VIA, {\it i.e.} neglecting soft-gluon emissions and the creation of mesonic intermediate states between initial and final mesons). The impact of the connected diagrams is of two kinds: for the operators ${\cal O}_1, {\cal O}_2, {\cal O}_3, {\cal O}_4$ and ${\cal O}_4^\prime$, we get a significant improvement on the bounds on the corresponding Wilson coefficients (but for $C_1^{XYZ}$ and $C_3^{XXZ}$, for which the improvement is rather small); or, for ${\cal O}_5$ and ${\cal O}_5^\prime$, we find that the bounds are a little looser than after the perturbative running (but still tighter than the tree-level ones). These behaviours can be easily traced back to the mixing induced by the inclusion of connected diagrams: for the first set of operators, we can see that operators with loose bounds (such as ${\cal O}_2$) get mixed
with operators with a tighter bound (in the case of ${\cal O}_2$, with ${\cal O}_1$). As a consequence, tighter constraints apply and 
we get that $C^{XXX}_2(2 \, {\rm GeV})$ indeed go from $2.8 \times 10^4$ to $4.5 \times 10^2$ ({\it i.e.} a reduction of a factor 62!), 
or $C^{XYZ}_3(2 \, {\rm GeV})$ go from $2.1 \times 10^4$ to $8.5 \times 10^2$ ({\it i.e.} a reduction of a factor 25!). On the other hand, ${\cal O}_5$ and ${\cal O}_5^\prime$ mix between themselves with a negative contribution of the connected diagram to the disconnected one, thus making the bound looser by ${\cal O}(1/N_c)$. 

When we move to Tab.~\ref{tab:results2}, we can see that
for two muons in the final state the situation is not much different: 
also in this case the biggest improvements with respect to the 
one-loop RG-evolved bounds can be found for $C_2^{XXZ}$, that
goes from $4.2 \times 10^4$ down to $6.7 \times 10^2$, and 
for $C_3^{XYZ}$, that goes from $3.4 \times 10^4$ down to $1.3 \times 10^3$ (in both cases, the improvement is indentical to that found for
$e e$ in the final state). 

\begin{table}[ht]
\centering
\begin{tabular}{|c|c|c|c|}
\hline
&&& \\
\textbf{Coefficient} & \textbf{Tree} & \textbf{RGE-Imp} \cite{Gonzalez:2023him} & \textbf{RGE + connected diag.} \\
&&& \\
\hline
\hline
$C_{1}^{XXZ}$     & $3.7\times 10^2$ & $2.2\times 10^2$ & $1.8\times 10^2$ \\
\hline
$C_{1}^{XYZ}$     & $3.7\times 10^2$ & $1.6\times 10^2$ & $1.6\times 10^2$ \\
\hline
$C_{2}^{XXZ}$     & --            & $4.2\times 10^4$ & $6.7\times 10^2$ \\
\hline 
$C_{3}^{XXZ}$     & $3.1\times 10^4$ & $3.8\times 10^4$ & $3.2\times 10^4$ \\
\hline
$C_{3}^{XYZ}$     & $3.1\times 10^4$ & $3.4\times 10^4$ & $1.3\times 10^3$   \\
\hline
$C_{4}^{XXZ}$     & --            & $5.6\times 10^4$ & $1.8\times 10^4$ \\
\hline
$C_{4}^{\prime XXZ}$    & --            & $6.3\times 10^4$ & $2.\times 10^4$ \\
\hline
$C_{4}^{XYZ}$     & --            & $2.5\times 10^5$ & $2.3\times 10^4$ \\
\hline
$C_{4}^{\prime XYZ}$    & --            & $2.3\times 10^5$ & $2.1\times 10^4 $ \\
\hline
$C_{5}^{XXZ}$     & $ 4.3\times 10^3$ & $2.8\times 10^3$ & $3.1\times 10^3$             \\
\hline
$C_{5}^{\prime XXZ}$    & $3.8\times 10^3$ & $2.5\times 10^3$ & $2.7\times 10^3$             \\
\hline
$C_{5}^{XYZ}$     & $4.3\times 10^3$ & $3.3\times 10^3$ & $3.6\times 10^3$ \\
\hline
$C_{5}^{\prime XYZ}$    & $3.8\times 10^3$ & $3.0\times 10^3$ & $3.2\times 10^3$ \\
\hline
\hline
\end{tabular}
\caption{\it Bounds on Wilson coefficients at $\mu = 2$ GeV 
from the process $K^+\to \pi^-\mu^+\mu^+$: at tree level, including RGE, and with the effect of connected diagrams.
Notice that in this case the bound apply to all lepton bilinear chiralities.
}
\label{tab:results2}
\end{table}
If we eventually focus on Tab.~\ref{tab:results3}, where our
results for two different lepton in the final state are shown, 
we see that for the operators ${\cal O}_1 - {\cal O}_5$ there
are no big differences with respect to the case of two identical leptons. On the other hand, for the new set of operators that
are only present in this case, ${\cal O}_6, {\cal O}_7$ and ${\cal O}_8$, we find no improvement for the bounds on the Wilson 
coefficients $C_6$, whilst at the same we get a significant
change in the bounds for $C_7^{XYY}, C_7^{\prime \, XYX}$, 
for which we go from $5.4 \times 10^5$ for the one-loop RG-evolved coefficients down to $7.3 \times 10^4$ for the full VIA result
(a reduction of a factor 7). Notice that, as it was the case
for both the tree-level and the RGE-improved results, we are
not able to constrain the Wilson coefficient of the operator ${\cal O}_8$.

\begin{table}[h]
\centering
\begin{tabular}{|c|c|c|c|}
\hline
&&& \\
\textbf{Coefficient} & \textbf{Tree} & \textbf{RGE-Imp} \cite{Gonzalez:2023him} & \textbf{RGE + connected diag.} \\
&&& \\
\hline
\hline
$C_{1}^{XXZ}$     & $2.0\times 10^2$ & $1.2\times 10^2$ & $9.1\times 10^1$\\
\hline
$C_{1}^{XYZ}$     & $2.0\times 10^2$ & $8.3\times 10^1$ & $8.3\times 10^1$ \\
\hline
$C_{2}^{XXX}$     & --            & $2.2\times 10^4$ & $3.6\times 10^2$ \\
\hline
$C_{3}^{XXZ}$     & $1.5\times 10^4$ & $1.9\times 10^4$ & $1.6\times 10^4$ \\
\hline
$C_{3}^{XYZ}$     & $1.5\times 10^4$ & $1.7\times 10^4$ & $6.7\times 10^2$             \\
\hline
$C_{4}^{XXZ}$     & --            & $2.6\times 10^4$ & $7.9\times 10^3$ \\
\hline
$C_{4}^{\prime XXZ}$    & --            & $2.9\times 10^4$ & $8.9\times 10^3$ \\
\hline
$C_{4}^{XYZ}$     & --            & $1.2\times 10^5$ & $1.1\times 10^4$ \\
\hline
$C_{4}^{\prime XYZ}$    & --            & $1.0\times 10^5$ & $9.4\times 10^3$ \\
\hline
$C_{5}^{XXZ}$     & $2.0\times 10^3$ & $1.3\times 10^3$ & $1.4\times 10^3$             \\
\hline
$C_{5}^{\prime XXZ}$    & $1.8\times 10^3$ & $1.2\times 10^3$ & $1.3\times 10^3$             \\
\hline
$C_{5}^{XYZ}$     & $2.0\times 10^3$ & $1.5\times 10^3$ & $1.7\times 10^3$ \\
\hline
$C_{5}^{\prime XYZ}$    & $1.8\times 10^3$ & $1.4\times 10^3$ & $1.5\times 10^3$ \\
\hline
$C_{6}^{XXX}$    & $3.2\times 10^4$ & $3.5\times 10^4$ & $3.0\times 10^4$ \\
\hline
$C_{6}^{XXY}$    & $3.2\times 10^4$ & $2.8\times 10^4$ & $3.4\times 10^4$ \\
\hline
$C_{6}^{XYX}$    & $3.2\times 10^4$ &  $3.4\times 10^4$ &  $3.4\times 10^4$ \\
\hline
$C_{6}^{XYY}$    & $3.2\times 10^4$ &  $3.4\times 10^4$ &  $3.4\times 10^4$ \\
\hline
$C_{7}^{XXX}$    & --              & --    & -- \\
\hline
$C_{7}^{XYY}$    & --              & $5.4\times 10^5$ & $7.3\times 10^4$ \\
\hline
$C_{7}^{\prime XXX}$    & --              & --  & -- \\
\hline
$C_{7}^{\prime XYX}$    & --              & $5.4\times 10^5$ & $7.3\times 10^4$ \\
\hline
$C_{8}^{XXX}$    & --              & --  & -- \\
\hline
\hline
\end{tabular}
\caption{\it Bounds on Wilson coefficients from the process $K^+\to \pi^-e^+\mu^+$ at $\mu = 2$ GeV: at tree level, including RGE, and with the effect of connected diagrams. Notice that for different final leptons, 
the bounds on the Wilson coefficients $C_1,C_3,C_4,C_4^\prime,C_5$ and $C_5^\prime$ do not depend on the chirality of the lepton bilinear, whereas for $C_2, C_6, C_7, C_7^\prime$ and $C_8$ they do.
}
\label{tab:results3}
\end{table}

\section{Conclusions}
\label{sec:concl}

In this paper we have considered lepton number violating meson 
decays. Building on previous work \cite{Gonzalez:2023him}, which
included QCD one-loop corrections as well as Renormalization Group
(RG) running in the matching of the $d=9$ LNV operators describing
these decays, we have calculated, for the first time in the literature,
${\cal O } (1/N_c)$ corrections to the hadronic matrix element.

Previous works on LNV meson decays have all estimated the hadronic matrix element by considering the disconnected diagram in the
so-called {\it Vacuum Insertion Approximation}, only. 
We first have discussed in detail, how this procedure describes
the leading term of the hadronic matrix element in the limit $N_c \to
\infty$. Connected diagrams, previously neglected, should be also included to have a consistent VIA estimate, though. These diagrams, 
after Fierz re-arrangement of spinor and color indices,
give ${\cal O } (1/N_c)$ corrections to the previous estimates
and induce mixing between different operators, 
which has to be re-diagonalized in order to
extract limits on Wilson coefficients.

Naive power counting in $(1/N_c)$ would lead to the expectation that
the inclusion of the connected ${\cal O } (1/N_c)$ diagrams, that we
have considered in this work, would change limits on the Wilson
coefficients by a factor roughly given as $(1+1/N_c$), i.e. a naive 30
\% effect. Our main numerical results, shown in tables \ref{tab:results1}, \ref{tab:results2} and \ref{tab:results3},
however, show that this is not always the case.  In some particular
cases bounds on Wilson coefficients change by nearly two orders of
magnitude. This can be traced back to the fact that the naive
expectation neglects that after the inclusion of connected diagrams
one needs to re-diagonalize the operator basis.  The effect is similar
to that found when the tree-level bounds are improved by taking into
account RG running. Operator mixing can lead, in some cases, to bounds
on Wilson coefficients, which were completely absent at tree-level.
That long-range contributions to the hadronic matrix elements can have
such an important effect on the Wilson coefficients has, to the best
of our knowledge, not been discussed in the literature before for
$\Delta L = 2$ operators. 

We should, however, also add a disclaimer here. While our calculation gives improved bounds for some Wilson coefficients, these bounds still do not provide competitive limits. This fact is easy to understand. Any of the $d=9$ $\Delta L = 2$ operators we consider can be generated from integrating out some beyond-the-SM particles in the UV. We can convert the bounds on the Wilson coefficients into a bound on the (geometric) mean mass of those BSM states. With even the best limits in our tables giving numbers much larger than one, the constraint on the mean mass will be below the mass of the W-boson. (Charged) BSM particles with such low masses should have already been found by LEP. Thus, in order for the lepton number violating meson and tau decays to become competitive the experimental limits need to be improved by at least four order of magnitudes.

Improving the experimental limits by such large factors seems a difficult task, but not impossible. The HIKE experiment at CERN \cite{HIKE:2023ext} was supposed to produce roughly $8\times 10^{13}$ $K^+$ mesons in the four years of running in its phase-I. And, according to the experimental collaboration, this would have allowed to probe exotic LNV decay modes of the kaon down to branching ratios of order $10^{-13}$. The beam dump of the DUNE experiment \cite{DUNE:2020lwj}, on the other hand, is estimated to produce $K^+$'s in excess of  $5 \times 10^{21}$ in 10 years of running \cite{Krasnov:2019kdc,Gunther:2023vmz}. One could expect, therefore, that installing a HIKE like detector in the decay tunnel of DUNE's beam dump could probe rare kaon decays several orders of magnitude better than HIKE. Of course, no such proposal exists to the best of our knowledge.

We close this paper by reiterating that the inclusion of the connected
${\cal O } (1/N_c)$ diagrams is just the beginning of a full,
non-perturbative computation of the long-range QCD effects in the
matching of the LNV operators. A complete realistic calculation can
only be performed using lattice field theory techniques. However,
connected diagrams represent the leading order in the $1/N_c$
expansion of the QCD non-perturbative effects and thus we are fairly
confident that our work presents a valid first step towards a complete
calculation of the long-range part of the matrix elements.

\newpage
\subsection*{Acknowledgements}\label{sec:acknowledgements}

AD is supported by the EU H2020 research and innovation programme
under the MSC grant agreement 860881-HIDDeN and the Staff Exchange
grant agreement 101086085 ASYMMETRY, by the Spanish Ministerio de
Ciencia e Innovación project PID2020-113644GB-I00, and by the
Generalitat Valenciana through the grant CIPROM/2022/69.  M.H. is
supported by spanish grant PID2023-147306NB-I00 and CIPROM/2021/054
(Generalitat Valenciana).  Both AD and MH acknowledge support by
CEX2023-001292-S (MCIU/AEI/10.13039/501100011033).
MG acknowledges support from Centro
de F\'isica Te\'orica de Valpara\'iso (CEFITeV) and project PFE UVA22991/PUENTE.
NN acknowledges support from ANID (Chile) FONDECYT Iniciaci\'on
Grant No. 11230879.

\begin{appendices}

\section{Anomalous dimension matrices}
\label{app:ADM}

We present in this Appendix the anomalous dimensions for the operators
in eqs.~(\ref{eq:relevantoperators})
and (\ref{eq:relevantoperators2}), computed in dimensional regularization in the $\overline{\mathrm{MS}}$ scheme at one-loop. 

For identical or different final leptons we have:\footnote{Notice that these matrices are not identical to those in ref. \cite{Gonzalez:2023him}, since in this work we are using a slightly different convention for the operator basis.}  

\begin{align}
\left \{
\begin{array}{l}
\hat{\gamma}^{XXZ}_{(12)}
=-2\left(
\begin{array}{cc}
6 C_F-3& \frac{N-2}{4 N}\\
-\frac{12(2+N)}{N}&-3-2 C_F
\end{array}
\right),\\
\\
\hat{\gamma}_{(13)}^{XYZ}
=-2\left(
\begin{array}{cc}
6 C_F&0\\
-6& -\frac{3}{N} 
\end{array}
\right), \\
\\
\hat{\gamma}_{(3)}^{XXZ}=-2\left(-3+\frac{3}{N}\right), \\
\\
\hat{\gamma}_{(4^{\prime}45^{\prime}5)}^{XXZ}
=-2\left(
\begin{array}{cccc}
-C_F & -\frac{3}{2} & - \frac{3i}{N} & - \frac{3i}{2}\\
-\frac{3}{2} & -C_F & - \frac{3 i}{2} & - \frac{3 i}{N}\\
 \frac{i}{N} & - \frac{i}{2} & 3 C_F & -\frac{3}{2}\\
- \frac{i}{2} &  \frac{i}{N} & -\frac{3}{2} & 3 C_F
\end{array}\right),\\
\\
\hat{\gamma}_{(4'5')}^{XYZ}=
-2\left(
\begin{array}{cc}
-\frac{3}{2}-C_F & \frac{3i(N+2)}{2N} \\
\frac{i(N-2)}{2N} & -\frac{3}{2}+3C_F 
\end{array}\right),
\end{array}
\right .
\end{align}

On the other hand, for different final leptons, we also have: 

\begin{align}
\left \{
\begin{array}{l}
\hat{\gamma}_{(6)}^{XXX} =-2 \left(-1 + \frac{1}{N}\right), \\
\\
\hat{\gamma}_{(6)}^{XXY} =-2 \left(1+\frac{1}{N}\right), \\
\\
\hat{\gamma}_{(67')}^{XYX} =-2 \left(
\begin{array}{cc}
 -\frac{1}{N} & -\frac{i}{2} \\
 0 & 2 C_F \\
\end{array}
\right), \\
\\
\hat{\gamma} _{(7'78)}^{XXX} =-2 \left(
\begin{array}{ccc}
 2 C_F-\frac{1}{N}-2 & -\frac{1}{N} & \frac{i(8+N)}{8 N} \\
 -\frac{1}{N} & 2 C_F-\frac{1}{N}-2 &  -\frac{i(8+N)}{8 N}\\
 0 & 0 & -2 C_F
\end{array}
\right).
\end{array}
\right .
\end{align}

\section{$\mu$-evolution matrices}\label{app:umatrices}

For completeness, in this Appendix we list the evolution matrices 
$\hat U(\mu,\Lambda)$, relevant for calculating the values in the 
third column (``RGE-Imp'') in Tabs.~\ref{tab:results1}, \ref{tab:results2} and \ref{tab:results3}.
These matrices were calculated in Ref.~\cite{Gonzalez:2023him} 
from the anomalous dimension matrix in App.~\ref{app:ADM}, for $\mu=2\,\mbox{GeV}$ and $\Lambda=m_Z$. 

For identical or different final leptons we have:\footnote{Notice that these matrices are not identical to those in Ref.~\cite{Gonzalez:2023him}, since in this work we are using a slightly different convention for the operator basis and we compute them at $\mu = 2$ GeV.} 

\begin{align}
\left \{
\begin{array}{l}
\hat U_{(12)}^{XXZ} = 
\left(
\begin{array}{cc}
 1.68 & 0.01 \\
 -2.12 & 0.55 \\
\end{array}
\right) \, , \\
\\
\hat U_{(13)}^{XYZ}  = 
\left(
\begin{array}{cc}
 2.31 & 0. \\
 -0.94 & 0.90 \\
\end{array}
\right) \, , \\
\\
\hat U_{(3)}^{XXZ}(iji\ell)=0.81 \, , \\
\\
\hat U_{(4'45'5)}^{XXZ}  =
\left(
\begin{array}{cccc}
 0.88 & -0.14 &  -0.09 i &  -0.17 i \\
 -0.14 & 0.88 &  -0.17 i &  -0.09 i \\
 0.05 i & -0.07 i & 1.54 & -0.24 \\
  -0.07 i & 0.05 i & -0.24 & 1.54 \\
\end{array}
\right) \, , \\
\\
\hat U_{(4'5')}^{XYZ} =
\left(
\begin{array}{cc}
 0.74 &  0.26 i \\
 0.02 i & 1.30 \\
\end{array}
\right) \, .
\end{array}
\right .
\end{align}

On the other hand, for different final leptons we also have:
\begin{align}
\left \{
\begin{array}{l}
\hat U_{(6)}^{XXX} =0.93 \, , \\
\\
\hat U_{(6)}^{XXY} =1.15 \, , \\
\\
\hat U_{(67')}^{XYX} =
\left(
\begin{array}{cc}
 0.97 &  -0.06i  \\
 0  & 1.32 \\
\end{array}
\right) \, , \\
\\
\hat U_{(7'78)}^{XXX} =
\left(
\begin{array}{ccc}
 1.04  & -0.04 & 0.04 i \\
 -0.04  & 1.04 &- 0.04 i \\
 0 & 0 & 0.76
\end{array}
\right) \, . 
\end{array}
\right .
\end{align}

\section{Fierz identities}
\label{app:fierz}

We write in this Appendix the Fierz identities used in Sect.~\ref{sec:connected}. They have been obtained in Minkowski space-time. The chiral projectors are defined as follow:
\begin{equation}
   P_{L,R} = \frac{1}{2} \, (1 \mp \gamma_5)
\end{equation} 
and the tensor Dirac matrix is:
\begin{equation}
    \sigma_{\mu\nu} = \frac{i}{2} \, [\gamma_{\mu},\gamma_{\nu}] \, .
\end{equation}
No specific convention is needed for $\gamma_5$. We list below the relevant Fierz identities needed to rearrange spinor indices of the four-fermion operators in eqs.~(\ref{eq:relevantoperators}) and (\ref{eq:relevantoperators2}).

\begin{enumerate}
\item ${\cal O}_1^{XYZ} = 8 \left ( \bar u^A P_X d^B \right ) \otimes \left (  \bar u^C P_Y s^D \right ) \otimes j^Z \otimes 
\left ( \delta_{AB} \, \delta_{CD} \right ) $ \\
\begin{eqnarray}
    \left ( \bar a^A_\alpha \, P_X^{\alpha \beta} b^B_\beta \right ) \otimes
    \left ( \bar c^C_\gamma \, P_X^{\gamma \delta} d^D_\delta \right)  &=&
    - \frac{1}{2}  \left (\bar a^A_\alpha \, P_X^{\alpha \delta} d^D_\delta \right ) \otimes 
    \left (\bar c^C_\gamma \, P_X^{\gamma \beta} b^B_\beta \right )  \\
    &&- \frac{1}{8}  \left ( \bar a^A_\alpha  \, \left [ \sigma_{\mu\nu} P_X \right ]^{\alpha \delta} d^D_\delta \right ) \otimes 
    \left ( \bar c^C_\gamma  \, \left [ \sigma^{\mu\nu} \, P_X \right ]^{\gamma \beta} b^B_\beta \right ) \, , \nonumber \\
    \nonumber \\
    \left (\bar a^A_\alpha \, P_X^{\alpha \beta} b^B_\beta \right )\otimes \left (\bar c^C_\gamma \, P_Y^{\gamma \delta} d^D_\delta \right ) 
    & = & - \frac{1}{2}  \left (\bar a^A_\alpha \, \left [ \gamma_{\mu} \, P_Y \right ]^{\alpha \delta} d^D_\delta \right ) \otimes 
                                 \left ( \bar c^C_\gamma \left [ \gamma^{\mu} \, P_X \right ]^{\gamma \beta} b^B_\beta \right ) \, ,
\end{eqnarray}
where $P_X = P_{L,R}$ and $P_Y = P_{R, L}$. In order to be extremely clear, we have explicitly written color indices ($A, B, C, D$) and spinor indices $(\alpha, \beta, \gamma, \delta)$
of the four-fermion operator. It can be seen that, whereas spinor indices of the Dirac matrices on the r.h.s. are correctly contracted over the spinor indices of the corresponding 
fermion fields, color indices do not match. In Sect.~\ref{sec:connected}, the rearrangement of color indices will be performed explicitly so to relate the connected diagrams
with the disconnected ones. 
\item ${\cal O}_2^{XXX} = 8 \left ( \bar u^A \sigma_{\mu\nu} \, P_X d^B \right ) \otimes \left (  \bar u^C \sigma^{\mu\nu} \, P_X s^D \right ) \otimes j^X \otimes \left ( \delta_{AB} \, \delta_{CD} \right ) $ \\
\begin{eqnarray}
    \left ( \bar a^A_\alpha \, \left [ \sigma_{\mu\nu} \, P_X \right ]^{\alpha \beta} b^B_\beta \right ) \otimes
    \left ( \bar c^C_\gamma  \, \left [ \sigma^{\mu\nu} \, P_X \right ]^{\gamma \delta} d^D_\delta \right)  &=&  
    - 6  \left (\bar a^A_\alpha  \, P_X^{\alpha \delta} d^D_\delta \right ) \otimes 
    \left (\bar c^C_\gamma \, P_X^{\gamma \beta} b^B_\beta \right ) \\
    &&+ \frac{1}{2}  \left ( \bar a^A_\alpha  \, \left [ \sigma_{\mu\nu} P_X \right ]^{\alpha \delta} d^D_\delta \right ) \otimes 
    \left ( \bar c^C_\gamma  \, \left [ \sigma^{\mu\nu} \, P_X \right ]^{\gamma \beta} b^B_\beta \right ) \, , \nonumber
\end{eqnarray}
where $P_X = P_{L,R}$. 
\item ${\cal O}_3^{XYZ} = 8 \left ( \bar u^A \gamma^\mu P_X d^B \right ) \otimes \left ( \bar u^C \gamma_\mu P_Y s^D \right ) \otimes j^Z 
\otimes \left ( \delta_{AB} \, \delta_{CD} \right ) $ \\
\begin{eqnarray}
   \left (\bar a^A_\alpha \left [ \gamma_{\mu} \, P_X \right ]^{\alpha \beta} b^B_\beta \right ) \otimes 
   \left ( \bar c^C_\gamma  \left [ \gamma^{\mu} \, P_X \right ]^{\gamma \delta} d^D_\delta \right )&=&
   \left (\bar a^A_\alpha  \left [\gamma_{\mu} \, P_X \right ]^{\alpha \delta} d^D_\delta \right )\otimes 
   \left ( \bar c^C_\gamma  \left [ \gamma^{\mu} \, P_X \right ]^{\gamma \beta} b^B_\beta \right ) \, ,  \nonumber \\
    \\
    \left (\bar a^A_\alpha  \left [ \gamma_{\mu} \, P_X \right ]^{\alpha \beta} b^B_\beta \right ) \otimes 
    \left ( \bar c^C_\gamma  \left [ \gamma_{\mu} \, P_Y \right ]^{\gamma \delta} d^D_\delta \right ) &=&
    -2 \left ( \bar a^A_\alpha  \, P_Y^{\alpha \delta} d^D_\delta \right ) \otimes 
         \left (\bar c^C_\gamma  P_X^{\gamma \beta} b^B_\beta \right ) \, ,
\end{eqnarray}
where $P_X = P_{L,R}$ and $P_Y = P_{R, L}$.
\item ${\cal O}_4^{XYZ} = 8 \left ( \bar u \gamma^\mu P_X d \right ) \otimes \left ( \bar u \sigma_\mu^\nu \, P_Y s \right ) \otimes j_\nu^Z \otimes \left ( \delta_{AB} \, \delta_{CD} \right ) $ \\
\begin{eqnarray}
   \left (\bar a^A_\alpha \left [ \gamma_{\mu} \, P_X \right ]^{\alpha \beta} b^B_\beta \right ) \otimes  
   \left ( \bar c^C_\gamma  \left [ \sigma^{\mu\nu} \, P_X \right ]^{\gamma \delta} d^D_\delta \right ) & = &
 \frac{1}{2} \left (\bar a^A_\alpha  \left [\gamma_{\mu} \, P_X \right ]^{\alpha \delta} d^D_\delta \right ) \, 
 \otimes 
   \left ( \bar c^C_\gamma  \left [ \sigma^{\mu\nu} \, P_X \right ]^{\gamma \beta} b^B_\beta \right ) \nonumber  \\
    &&    - \frac{3 i}{2} \left (\bar a^A_\alpha  \left [ \gamma^\nu \, P_X \right ]^{\alpha \delta} d^D_\delta \right ) \otimes 
                   \left ( \bar c^C_\gamma P_X^{\gamma \beta} b^B_\beta \right ) \, ,  \nonumber \\
   \\
\nonumber     \\
    \left (\bar a^A_\alpha  \left [ \gamma_{\mu} \, P_X \right ]^{\alpha \beta} b^B_\beta \right ) \otimes 
    \left ( \bar c^C_\gamma  \left [ \sigma^{\mu\nu} \, P_Y \right ]^{\gamma \delta} d^D_\delta \right ) &=&
    \frac{1}{2} \left ( \bar a^A_\alpha  \, 
    \left [\sigma^{\mu\nu} \, P_Y \right ]^{\alpha \delta} d^D_\delta \right ) \,  
    \otimes 
         \left (\bar c^C_\gamma  
         \left [\gamma_\mu \, P_X \right ]^{\gamma \beta} b^B_\beta \right ) \nonumber \\
     && + \frac{3 i}{2} \left (\bar a^A_\alpha  P_Y^{\alpha \delta} d^D_\delta \right ) \otimes 
                   \left ( \bar c^C_\gamma   \left [ \gamma^\nu \, P_X \right ]^{\gamma \beta} b^B_\beta \right ) \, , \nonumber \\
\end{eqnarray}
where $P_X = P_{L,R}$ and $P_Y = P_{R, L}$.
\item ${\cal O}_4^{\prime \, XYZ} = 8 \left ( \bar u \sigma_\mu^\nu \, P_X d \right ) \otimes \left (\bar u \gamma^\mu P_Y s \right ) \otimes j^Z_\nu \otimes \left ( \delta_{AB} \, \delta_{CD} \right ) $ \\
These Fierz identities can be derived from those for ${\cal O}_4^{XYZ}$ by changing the quark flavours, $(\bar a,b) \leftrightarrow (\bar c,d)$.
\item ${\cal O}_5^{XYZ} = 8 \left ( \bar u \gamma^\mu P_X d \right ) \otimes \left ( \bar u P_Y s \right ) \otimes j^Z_\mu
\otimes \left ( \delta_{AB} \, \delta_{CD} \right )$ \\
\begin{eqnarray}
  \left (\bar a^A_\alpha \left [ \gamma_{\mu} \, P_X \right ]^{\alpha \beta} b^B_\beta \right )\otimes 
  \left (\bar c^C_\gamma P_X^{\gamma \delta} d^D_\delta \right )&=&
  -\frac{1}{2} \left (\bar a^A_\alpha  \left [ \gamma_{\mu} \, P_X \right ]^{\alpha \delta} d^D_\delta \right )\otimes
                   \left (\bar c^C_\gamma P_X^{\gamma \beta} b^B_\beta \right )  \\
  && + \frac{i}{2} \left (\bar a^A_\alpha  \left [ \gamma_\nu \, P_X \right ]^{\alpha \delta} d^D_\delta \right ) \otimes 
                   \left ( \bar c^C_\gamma  \left [\sigma^{\nu \mu} \, P_X \right ]^{\gamma \beta} b^B_\beta \right ) \, , \nonumber \\
   \left  (\bar a^A_\alpha  \left [ \gamma_{\mu} \, P_X \right ]^{\alpha \beta} b^B_\beta \right )\otimes 
   \left (\bar c^C_\gamma  P_Y^{\gamma \delta} d^D_\delta \right ) &=&
   -\frac{1}{2} \left (\bar a^A_\alpha  P_Y^{\alpha \delta} d^D_\delta \right )\otimes 
   \left (\bar c^C_\gamma  \left [\gamma_\mu \, P_X \right ]^{\gamma \beta} b^B_\beta \right) \\
&&   +\frac{i}{2} \left (\bar a^A_\alpha  \left [ \sigma^{\mu\nu} \, P_Y \right ] ^{\alpha \delta} d^D_\delta \right )\otimes 
   \left (\bar c^C_\gamma  \left [\gamma_\nu \, P_X \right ]^{\gamma \beta} b^B_\beta \right ) \, . \nonumber
\end{eqnarray}
\item ${\cal O}_5^{\prime \, XYZ} = 8 \left ( \bar u P_X d \right ) \otimes \left (\bar u \gamma^\mu P_Y s \right ) \otimes j^Z_\mu
\otimes \left ( \delta_{AB} \, \delta_{CD} \right ) $ \\
These Fierz identities can be derived from those for ${\cal O}_5^{XYZ}$ by changing the quark flavours, $(\bar a,b) \leftrightarrow (\bar c,d)$.
\item ${\cal O}_6^{XYZ} = 8 \left ( \bar u \gamma^\mu P_X d \right ) \otimes \left ( \bar u \gamma^\nu P_Y s \right ) \otimes j^Z_{\mu\nu}
\otimes \left ( \delta_{AB} \, \delta_{CD} \right ) $ \\
For this operator we have several choices, depending on the chirality of the lepton current:
\begin{eqnarray}
  \left ( \bar a^A_\alpha  \left [ \gamma_{\mu} \, P_X\right ]^{\alpha \beta} b^B_\beta \right )  
    & \otimes &
    \left (\bar c^C_\gamma  \left [ \gamma_{\nu} \, P_X \right ]^{\gamma \delta} d^D_\delta \right )  
    \otimes  
    j^{\mu\nu}_X =  \nonumber\\
    && -  \left (\bar a^A_\alpha  \left [ \gamma_{\mu} \, P_X \right ]^{\alpha \delta} d^D_\delta \right ) 
    \otimes
      \left ( \bar c^C_\gamma  \left [ \gamma_{\nu} \, P_X \right ]^{\gamma \beta} b^B_\beta \right )\otimes  
      j^{\mu\nu}_X \, , 
\end{eqnarray}
 \begin{eqnarray}
    \left ( \bar a^A_\alpha \left [ \gamma_{\mu} \, P_X\right ]^{\alpha \beta} b^B_\beta \right )
    & \otimes &
    \left (\bar c^C_\gamma \left [ \gamma_{\nu} \, P_X \right ]^{\gamma \delta} d^D_\delta \right ) 
    \otimes 
    j^{\mu\nu}_Y = \nonumber\\
    && \left  ( \bar a^A_\alpha  \left [ \gamma_{\mu} \, P_X \right ] ^{\alpha \delta} d^D_\delta \right )
    \otimes 
    \left ( \bar c^C_\gamma \left [ \gamma_{\nu} \, P_X \right ]^{\gamma \beta} b^B_\beta \right )
    \otimes 
    j^{\mu\nu}_Y \, ,
 \end{eqnarray}
 \begin{eqnarray}
  \left  ( \bar a^A_\alpha \left [ \gamma_{\mu} \, P_X \right ]^{\alpha \beta} b^B_\beta \right ) 
  & \otimes & 
  \left ( \bar c^C_\gamma  \left [ \gamma_{\nu} \, P_Y \right ]^{\gamma \delta} d^D_\delta \right )
  \otimes
  j^{\mu\nu}_X 
  =  \nonumber\\
  && -\frac{i}{2}
     \left ( \bar a^A_\alpha  P_Y^{\alpha \delta} d^D_\delta \right )
     \otimes 
     \left ( \bar c^C_\gamma \left [ \sigma_{\mu\nu}\, P_X \right ]^{\gamma \beta} b^B_\beta \right )
     \otimes 
     j^{\mu\nu}_X \, ,
  \end{eqnarray}
  \begin{eqnarray}
 \left  ( \bar a^A_\alpha  \left [ \gamma_{\mu} \, P_X \right ]^{\alpha \beta} b^B_\beta \right ) 
 & \otimes & 
        \left ( \bar c^C_\gamma  \left [ \gamma_{\nu} \, P_Y \right ]^{\gamma \delta} d^D_\delta \right )
        \otimes 
        j^{\mu\nu}_Y 
    = \nonumber\\
 && \frac{i}{2}
  \left  (\bar a^A_\alpha \left [ \sigma_{\mu\nu} \, P_Y \right ]^{\alpha \delta} d^D_\delta \right )
  \otimes
  \left (\bar c^C_\gamma  P_X^{\gamma \beta} b^B_\beta \right ) 
  \otimes 
  j^{\mu\nu}_Y \, .
  \end{eqnarray}
  \item ${\cal O}_7^{XZZ} = 8 \left ( \bar u  P_X d \right ) \otimes \left ( \bar u \sigma^{\mu \nu} \,  P_Z s \right ) \otimes j^Z_{\mu\nu} \otimes \left ( \delta_{AB} \, \delta_{CD} \right ) $ \\
  For this operator we have several choices, depending on the chirality of the lepton current:
  \begin{eqnarray}
  \left ( \bar a^A_\alpha P_X^{\alpha \beta} b^B_\beta \right )  & \otimes  &
    \left (\bar c^C_\gamma \left [ \sigma^{\mu\nu} \, P_X \right ]^{\gamma \delta} d^D_\delta \right )   \otimes   j_{\mu\nu}^X = \nonumber \\
   &&  -  \frac{1}{2} \left (\bar a^A_\alpha  \left [ \sigma^{\mu\nu} \, P_X \right ]^{\alpha \delta} d^D_\delta \right ) \otimes
      \left ( \bar c^C_\gamma  P_X^{\gamma \beta} b^B_\beta \right )\otimes  j_{\mu\nu}^X \nonumber \\
      && 
         -  \frac{1}{2} \left (\bar a^A_\alpha P_X^{\alpha \delta} d^D_\delta \right ) \otimes
      \left ( \bar c^C_\gamma   \left [ \sigma^{\mu\nu} \, P_X \right ]^{\gamma \beta} b^B_\beta \right )\otimes  j_{\mu\nu}^X \nonumber \\
      &&   
      +  \frac{i}{2} \left (\bar a^A_\alpha  \left [ \sigma_{\mu \rho}P_X \right ]^{\alpha \delta} d^D_\delta \right ) \otimes
      \left ( \bar c^C_\gamma   \left [ \sigma_\nu^\rho \, P_X \right ]^{\gamma \beta} b^B_\beta \right )\otimes  j_{\mu\nu}^X \, , \\
      \nonumber \\
      \left ( \bar a^A_\alpha  P_X^{\alpha \beta} b^B_\beta \right )  & \otimes  &
    \left (\bar c^C_\gamma  \left [ \sigma^{\mu\nu} \, P_Z \right ]^{\gamma \delta} d^D_\delta \right )   \otimes   j_{\mu\nu}^Z = \nonumber \\
   &&  2 i \, \left (\bar a^A_\alpha  \left [ \gamma^\mu \, P_Z \right ]^{\alpha \delta} d^D_\delta \right ) \otimes
      \left ( \bar c^C_\gamma  \left [ \gamma^\nu \, P_X \right ]^{\gamma \beta} b^B_\beta \right )\otimes  j_{\mu\nu}^Z    \, . \nonumber \\
 \end{eqnarray}
    \item ${\cal O}_7^{\prime \, XZZ} = 8 \left ( \bar u  \sigma^{\mu \nu} \,  P_X d \right ) \otimes \left ( \bar u  P_Z s \right ) \otimes j^Z_{\mu\nu} \otimes \left ( \delta_{AB} \, \delta_{CD} \right ) $ \\
    These Fierz identities can be derived from those for ${\cal O}_7^{XZZ}$ by changing the quark flavours, $(\bar a,b) \leftrightarrow (\bar c,d)$.
    \item ${\cal O}_8^{XXX} = 8 \left ( \bar u  \sigma^{\mu \rho} P_X d \right ) \otimes \left ( \bar u \sigma^\nu_\rho \,  P_X s \right ) \otimes j^X_{\mu\nu} \otimes \left ( \delta_{AB} \, \delta_{CD} \right )$ \\
    In this last case only the four-quark operator with chiral structure $XXX$ is non-vanishing:
    \begin{eqnarray}
      \left ( \bar a^A_\alpha \left [ \sigma^{\mu \rho} \, P_X \right ]^{\alpha \beta} b^B_\beta \right )  & \otimes  &
    \left (\bar c^C_\gamma  \left [ \sigma^\nu_\rho \, P_X \right ]^{\gamma \delta} d^D_\delta \right )   \otimes   j_{\mu\nu}^X = \nonumber \\
   &&    \frac{1}{8} \left (\bar a^A_\alpha  \left [ \sigma^{\mu\rho} \, P_X \right ]^{\alpha \delta} d^D_\delta \right ) \otimes
      \left ( \bar c^C_\gamma  \left [\sigma^\nu_\rho P_X \right ]^{\gamma \beta} b^B_\beta \right )\otimes  j_{\mu\nu}^X \nonumber \\
      && 
         + i \left (\bar a^A_\alpha  \left [ \sigma^{\mu\nu} \, P_X \right ]^{\alpha \delta} d^D_\delta \right ) \otimes
      \left ( \bar c^C_\gamma  P_X^{\gamma \beta} b^B_\beta \right )\otimes  j_{\mu\nu}^X \nonumber \\
      &&   
      - i \left (\bar a^A_\alpha  P_X^{\alpha \delta} d^D_\delta \right ) \otimes
      \left ( \bar c^C_\gamma   \left [ \sigma^{\mu\nu} \, P_X \right ]^{\gamma \beta} b^B_\beta \right )\otimes  j_{\mu\nu}^X \, .
\end{eqnarray}
  \end{enumerate}

\section{Sum of connected and disconnected diagrams in the VIA}
\label{app:sums}

We compute here the sum of connected and disconnected diagrams
in the VIA, using reorganization of spinor indices (according to the Fierz rules given in App.~\ref{app:fierz}) and reorganization of color indices following eq.~(\ref{eq:colorfierz}). 

We have: 
\begin{enumerate}
\item ${\cal O}_1^{XYZ}$\\
\begin{equation}
\label{eq:fierzconnectedO1XYZ}
\begin{array}{l}
8 \langle \pi^-  | \bar u P_X d  \, \bar u P_X s  | K^+ \rangle \otimes j^Z_{\ell \ell}  = 
8 \left \{ \left (1 - \frac{1}{4 N_c} \right ) \, \langle \pi^-  | \bar u P_X d  | 0 \rangle  \, \langle 0 |  \bar u P_X s  | K^+ \rangle \right . \\
\\
\qquad  \qquad \qquad \qquad  \qquad \qquad 
-  \left . \frac{1}{16 N_c}  \, \langle \pi^-  | \bar u \, \sigma_{\mu\nu} \, P_X d  | 0 \rangle  \, \langle 0 |  \bar u \, \sigma^{\mu\nu} \, P_X s  | K^+ \rangle  \right \} \otimes j^Z_{\ell \ell} \, 
+ \, \dots \, , \\
\\
8 \langle \pi^-  | \bar u P_X d  \, \bar u P_Y s  | K^+ \rangle \otimes j^Z_{\ell \ell}  = 
8 \left \{ 
\langle \pi^-  | \bar u  P_X d  | 0 \rangle  \, \langle 0 |  \bar u  P_Y s  | K^+ \rangle \right. \\
\\
\qquad  \qquad \qquad \qquad  \qquad \qquad 
- \left . \frac{1}{4 N_c}  
\langle \pi^-  | \bar u \gamma_\mu P_X d  | 0 \rangle  \, \langle 0 |  \bar u \gamma^\mu P_Y s  | K^+ \rangle \right \} \otimes j^Z_{\ell \ell}  \, + \, 
\dots \, , \\
\end{array}
\end{equation}
where the dots stand for the octet-octet contribution coming from the $t^a \otimes t^a$ term in the color-rearrangement rule of eq.~(\ref{eq:colorfierz}). This contribution (that is subleading in $1/N_c$) vanishes in the VIA, since the vacuum is a color singlet. 
As we can see, the operator $P_{L,R} \otimes P_{L,R}$ mixes with 
$\sigma_{\mu\nu} P_{L,R} \otimes \sigma^{\mu\nu}_{L,R}$ under Fierzing. 
Since in the VIA the matrix element of the tensor current vanishes, 
the operator $P_{L,R} \otimes P_{R,L}$ only mixes with the ${\cal O}_3^{XYZ}$ operator.
\item ${\cal O}_2^{XXX}$\\
\begin{equation}
\label{eq:fierzconnectedO2XXX}
\begin{array}{l}
8 \langle \pi^-  | \bar u \sigma^{\mu\nu} \, P_X d  \, \bar u \sigma_{\mu\nu} \, P_X s  | K^+ \rangle \otimes j^X_{\ell \ell}  = 
8 \left \{ \left (1 + \frac{1}{4 N_c} \right ) \, \langle \pi^-  | \bar u \sigma^{\mu\nu} \, P_X d  | 0 \rangle  \, \langle 0 |  \bar u \sigma_{\mu\nu} \, P_X s  | K^+ \rangle \right . \\
\\
\qquad  \qquad \qquad \qquad  \qquad \qquad 
-  \left . \frac{3}{N_c}  \, \langle \pi^-  | \bar u  P_X d  | 0 \rangle  \, \langle 0 |  \bar u P_X s  | K^+ \rangle  \right \} \otimes j^X_{\ell \ell} \, 
+ \, \dots \, .
\end{array}
\end{equation}
Notice that, once projected onto the vacuum, the leading contribution
proportional to the operator $\sigma_{\mu\nu} P_X \otimes \sigma^{\mu\nu} P_X$ vanishes in the VIA due to eq.~(\ref{eq:PCACtensor}). Therefore, the dominant contribution to the
matrix element of ${\cal O}_2$ is the $1/N_c$ term proportional 
to $P_X \otimes P_X$ that comes from the connected diagram.

\item {\bf ${\cal O}_3^{XYZ}$} \\
\begin{equation}
\label{eq:fierzconnectedO3XYZ}
\begin{array}{l}
8 \langle \pi^-  | \bar u \gamma^\mu P_X d  \, \bar u \gamma_\mu P_X s  | K^+ \rangle \otimes j^Z_{\ell \ell}  = 
8 \left \{ \left (1 + \frac{1}{ 2 N_c} \right ) \, \langle \pi^-  | \bar u \gamma^\mu P_X d  | 0 \rangle  \, \langle 0 |  \bar u \gamma_\mu P_X s  | K^+ \rangle  \right \} \otimes j^Z_{\ell \ell}  \, 
+ \,  \dots \, , \\
\\
8 \langle \pi^-  | \bar u \gamma^\mu P_X d  \, \bar u \gamma_\mu P_Y s  | K^+ \rangle \otimes j^Z_{\ell \ell}  = 
8 \left \{  \langle \pi^-  | \bar u \gamma^\mu P_X d  | 0 \rangle  \, \langle 0 |  \bar u \gamma_\mu P_Y s  | K^+ \rangle \right . \\
  \\
  \qquad \qquad  \qquad \qquad \qquad  \qquad \qquad 
  \left .   - \frac{1}{N_c} 
  \langle \pi^-  | \bar u P_X d  | 0 \rangle  \, \langle 0 |  \bar u P_Y s  | K^+ \rangle
  \right  \} \otimes j^Z_{\ell \ell}  \, + \, \dots \, . \\
\end{array}
\end{equation}
Notice that, as is well-known, four-fermion quark operators with a spinor structure as $\gamma_\mu P_{L,R} \otimes \gamma^{\mu} P_{ L, R}$ are self-Fierzing, and no mixing with other operators arise. The connected diagram just contributes a $1/N_c$ correction to the disconnected one. On the other hand, for $X \neq Y$ we get that $\gamma_\mu P_{L,R} \otimes \gamma^{\mu} P_{ R, L}$ mixes with ${\cal O}_1^{XYZ}$.

\item {\bf ${\cal O}_4^{XYZ} $ } \\
In this case, we have two options, for identical chiral projectors within the two quark currents, ${\cal O}_4^{XXZ}$, or different, ${\cal O}_4^{XYZ}$:
\begin{equation}
\label{eq:fierzconnectedO4XYZ1}
\begin{array}{l}
8 \langle \pi^-  | \bar u \, \gamma^\mu \, P_X d  \, \bar u  \sigma_\mu^\nu \, P_X s  | K^+ \rangle \otimes j^Z_{\ell \ell, \nu} = 
8 \left \{ 
\langle \pi^-  | \bar u \, \gamma^\mu P_X d  | 0 \rangle  \, \langle 0 |  \bar u \sigma_\mu^\nu \, P_X s  | K^+ \rangle \right . \\
\\
\qquad  \qquad \qquad \qquad  \qquad \qquad 
+   \frac{1}{4 N_c}  \, \langle \pi^-  | \bar u \sigma_\mu^\nu \, P_X d  | 0 \rangle  \, \langle 0 |  \bar u \, \gamma^\mu \, P_X s  | K^+ \rangle \\
\\
 \qquad  \qquad \qquad \qquad  \qquad \qquad 
 - \left .  \frac{3 i}{4 N_c}  \, \langle \pi^-  | \bar u \, P_X d  | 0 \rangle  \, \langle 0 |  \bar u \, \gamma_\nu \, P_X s  | K^+ \rangle  \right \} 
 \otimes j^Z_{\ell \ell, \nu} 
 \, + \, \dots \, , \\
 \\
 8 \langle \pi^-  | \bar u \, \gamma^\mu \, P_X d  \, \bar u  \sigma_\mu^\nu \, P_Y s  | K^+ \rangle \otimes j^Z_{\ell \ell, \nu}  = 
8 \left \{  \left ( 1 + \frac{1}{4 N_c}  \right ) \,  \langle \pi^-  | \bar u \, \gamma^\mu P_X d  | 0 \rangle  \, \langle 0 |  \bar u \sigma_\mu^\nu \,  P_Y s  | K^+ \rangle \right . \\
\\
 \qquad  \qquad \qquad \qquad  \qquad \qquad 
 + \left .  \frac{3 i}{4 N_c}  \, \langle \pi^-  | \bar u \, \gamma_\nu \, P_X d  | 0 \rangle  \, \langle 0 |  \bar u \, P_Y s  | K^+ \rangle  \right \} 
 \otimes j^Z_{\ell \ell, \nu} \, + \, 
\dots \, . \\
\end{array}
\end{equation}
Also in this case, as it was for ${\cal O}_2$, the dominant contribution in the VIA comes from the connected diagram, since the leading  disconnected one vanishes due to eq.~(\ref{eq:PCACtensor}). 

\item {\bf ${\cal O}_4^{\prime XYZ} $ } \\
We have:
\begin{equation}
\label{eq:fierzconnectedO4XYZ2}
\begin{array}{l}
8 \langle \pi^-  | \bar u \, \sigma_\mu^\nu \, P_X d  \, \bar u   \gamma^\mu \, P_X s  | K^+ \rangle \otimes j^Z_{\ell \ell, \nu} = 
8 \left \{ 
\langle \pi^-  | \bar u \,  \sigma_\mu^\nu \, P_X d  | 0 \rangle  \, \langle 0 |  \bar u \gamma^\mu \, P_X s  | K^+ \rangle \right . \\
\\
\qquad  \qquad \qquad \qquad  \qquad \qquad 
+   \frac{1}{4 N_c}  \, \langle \pi^-  | \bar u \gamma^\mu \, P_X d  | 0 \rangle  \, \langle 0 |  \bar u \, \sigma_\mu^\nu \,  P_X s  | K^+ \rangle \\
\\
 \qquad  \qquad \qquad \qquad  \qquad \qquad 
 - \left .  \frac{3 i}{4 N_c}  \, \langle \pi^-  | \bar u \,  \gamma_\nu \, P_X d  | 0 \rangle  \, \langle 0 |  \bar u \, P_X s  | K^+ \rangle  \right \} 
 \otimes j^Z_{\ell \ell, \nu} 
 \, + \, \dots \, , \\
 \\
 8 \langle \pi^-  | \bar u \, \sigma_\mu^\nu \, P_X d  \, \bar u  \, \gamma^\mu \,  P_Y s  | K^+ \rangle \otimes j^Z_{\ell \ell, \nu}  = 
8 \left \{  \left ( 1 + \frac{1}{4 N_c}  \right ) \,  \langle \pi^-  | \bar u \,  \sigma_\mu^\nu \,  P_X d  | 0 \rangle  \, \langle 0 |  \bar u \, \gamma^\mu  \, P_Y s  | K^+ \rangle \right . \\
\\
 \qquad  \qquad \qquad \qquad  \qquad \qquad 
 + \left .  \frac{3 i}{4 N_c}  \, \langle \pi^-  | \bar u  \, P_X d  | 0 \rangle  \, \langle 0 |  \bar u \, \gamma_\nu \, P_Y s  | K^+ \rangle  \right \} 
 \otimes j^Z_{\ell \ell, \nu} \, + \, 
\dots \, . \\
\end{array}
\end{equation}

\item {\bf ${\cal O}_5^{XYZ} $ } \\
Also in this case, we have two options:
\begin{equation}
\label{eq:fierzconnectedO5XYZ1}
\begin{array}{l}
8 \langle \pi^-  | \bar u \, \gamma^\mu P_X d  \, \bar u  P_X s  | K^+ \rangle \otimes j^Z_{\ell \ell, \mu} = 
8 \left \{ \langle \pi^-  | \bar u \, \gamma^\mu P_X d  | 0 \rangle  \, \langle 0 |  \bar u \ P_X s  | K^+ \rangle \right . \\
\\
\qquad  \qquad \qquad \qquad  \qquad \qquad -   \frac{1}{4 N_c}  \, \langle \pi^-  | \bar u P_X d  | 0 \rangle  \, \langle 0 |  \bar u \, \gamma^\mu P_X s  | K^+ \rangle \\
\\
 \qquad  \qquad \qquad \qquad  \qquad \qquad 
 + \left .  \frac{i}{4 N_c}  \, \langle \pi^-  | \bar u \, \sigma^{\alpha\mu} \, P_X d  | 0 \rangle  \, \langle 0 |  \bar u \, \gamma_\alpha \, P_X s  | K^+ \rangle  \right \} 
 \otimes j^Z_{\ell \ell, \mu} 
 \, + \, \dots \, , \\
 \\
 8 \langle \pi^-  | \bar u \, \gamma^\mu P_X d  \, \bar u  P_Y s  | K^+ \rangle \otimes j^Z_{\ell \ell, \mu}  = 
8 \left \{  \left ( 1 - \frac{1}{4 N_c}  \right ) \,  \langle \pi^-  | \bar u \, \gamma^\mu P_X d  | 0 \rangle  \, \langle 0 |  \bar u \ P_Y s  | K^+ \rangle \right . \\
\\
 \qquad  \qquad \qquad \qquad  \qquad \qquad 
 - \left .  \frac{i}{4 N_c}  \, \langle \pi^-  | \bar u \, \gamma_\alpha \, P_X d  | 0 \rangle  \, \langle 0 |  \bar u \, \sigma^{\alpha\mu} \, P_Y s  | K^+ \rangle  \right \} 
 \otimes j^Z_{\ell \ell, \mu} \, + \, 
\dots \, , \\
\end{array}
\end{equation}
We see that mixing arises with the tensor operator $\sigma P_{L,R}$. However, as the corresponding matrix element between a meson and the vacuum vanishes in the VIA, 
the only mixing that is induced by the Fierz-rearrangement is that between ${\cal O}_5$ and ${\cal O}_5^\prime$. 

\item {\bf ${\cal O}_5^{\prime XYZ}$} \\
\begin{equation}
\label{eq:fierzconnectedO5XYZ2}
\begin{array}{l}
8 \langle \pi^-  | \bar u  P_X d  \, \bar u \, \gamma^\mu P_X s  | K^+ \rangle \otimes j^Z_{\ell \ell, \mu}  = 
8 \left \{ \langle \pi^-  | \bar u P_X d | 0 \rangle  \, \langle 0 |  \bar u \, \gamma^\mu P_X s  | K^+ \rangle \right . \\
\\
\qquad  \qquad \qquad \qquad  \qquad \qquad 
-   \frac{1}{4 N_c}  \, \langle \pi^-  | \bar u \, \gamma^\mu \, P_X d  | 0 \rangle  \, \langle 0 |  \bar u P_X s  | K^+ \rangle \\
\\
 \qquad  \qquad \qquad \qquad  \qquad \qquad 
 + \left .  \frac{i}{4 N_c}  \, \langle \pi^-  | \bar u \, \gamma_\alpha \, P_X d  | 0 \rangle  \, \langle 0 |  \bar u \, \sigma^{\alpha \mu} \, P_X s  | K^+ \rangle  \right \} 
 \otimes j^Z_{\ell \ell, \mu} 
 \, + \, \dots \, , \\
\\
8 \langle \pi^-  | \bar u  P_X d  \, \bar u \, \gamma^\mu P_Y s  | K^+ \rangle \otimes j^Z_{\ell \ell, \mu}  = 
8 \left \{ \left ( 1 -  \frac{1}{4 N_c} \right ) \,  \langle \pi^-  | \bar u P_X d | 0 \rangle  \, \langle 0 |  \bar u \, \gamma^\mu P_Y s  | K^+ \rangle \right . \\
\\
 \qquad  \qquad \qquad \qquad  \qquad \qquad 
 + \left .  \frac{i}{4 N_c}  \, \langle \pi^-  | \bar u \, \sigma^{\mu \alpha} \, P_X d  | 0 \rangle  \, \langle 0 |  \bar u \, \gamma_\alpha \, P_Y s  | K^+ \rangle  \right \} 
 \otimes j^Z_{\ell \ell, \mu} 
 \, + \, \dots \, , \\
\end{array}
\end{equation}
We have the same results as for ${\cal O}_5^{XYZ}$.

\item {\bf ${\cal O}_6^{XYZ}$} \\
\begin{equation}
\label{eq:fierzconnectedO6XYZ1}
\begin{array}{l}
8 \langle \pi^-  | \bar u  \gamma^\mu P_X d  \, \bar u \, \gamma^\nu P_X s  | K^+ \rangle \otimes j^X_{\ell \ell , \mu\nu}  = 
8 \left ( 1 + \frac{1}{2 N_c} \right ) \, 
\langle \pi^-  | \bar u \gamma^\mu P_X d | 0 \rangle  \, \langle 0 |  \bar u \, \gamma^\nu P_X s  | K^+ \rangle \otimes j^X_{\ell \ell , \mu\nu} \, + \, 
\dots \, , \\
\\
8 \langle \pi^-  | \bar u  \gamma^\mu P_X d  \, \bar u \, \gamma^\nu P_X s  | K^+ \rangle \otimes j^Z_{\ell \ell , \mu\nu}  = 
8 \left ( 1 - \frac{1}{2 N_c} \right ) \, 
\langle \pi^-  | \bar u \gamma^\mu P_X d | 0 \rangle  \, \langle 0 |  \bar u \, \gamma^\nu P_X s  | K^+ \rangle \otimes j^Z_{\ell \ell , \mu\nu} 
\, + \, \dots \, , \\
\end{array}
\end{equation}
with $Z \neq X$ in the last equation. In order to get the second line of each Fierz identity we have used anti-commutation of 
Lorentz indices in the lepton current $j^Z_{\ell \ell , \mu\nu}$.

The last relation is: 
\begin{equation}
\label{eq:fierzconnectedO6XYZ2}
\begin{array}{l}
8 \langle \pi^-  | \bar u  \gamma^\mu P_X d  \, \bar u \, \gamma^\nu P_Y s  | K^+ \rangle \otimes j^X_{\ell \ell , \mu\nu}  = 
8 \left \{ \langle \pi^-  | \bar u \gamma^\mu P_X d | 0 \rangle  \, \langle 0 |  \bar u \, \gamma^\nu P_Y s  | K^+ \rangle \right . \\
\\
\qquad  \qquad \qquad \qquad  \qquad \qquad 
-  \left .  \frac{i}{4 N_c}  \, \langle \pi^-  | \bar u \, \sigma^{\mu \nu} \, P_X d  | 0 \rangle  \, \langle 0 |  \bar u P_Y s  | K^+ \rangle  \right \} \otimes j^X_{\ell \ell , \mu\nu} 
\, + \, \dots \, , \\
\\
8 \langle \pi^-  | \bar u  \gamma^\mu P_X d  \, \bar u \, \gamma^\nu P_Y s  | K^+ \rangle \otimes j^Y_{\ell \ell , \mu\nu}  = 
8 \left \{ \langle \pi^-  | \bar u \gamma^\mu P_X d | 0 \rangle  \, \langle 0 |  \bar u \, \gamma^\nu P_Y s  | K^+ \rangle \right . \\
\\
\qquad  \qquad \qquad \qquad  \qquad \qquad 
+   \left . \frac{i}{4 N_c}  \, \langle \pi^-  | \bar u P_X d  | 0 \rangle  \, \langle 0 |  \bar u \sigma^{\mu\nu} P_Y s  | K^+ \rangle  \right \} \otimes j^Y_{\ell \ell , \mu\nu} 
\, + \, \dots \, . \\
\end{array}
\end{equation}
\item {\bf ${\cal O}_7^{XZZ}$} \\
We have in this case:
\begin{equation}
\label{eq:fierzconnectedO7XZZ1}
\begin{array}{l}
8 \langle \pi^-  | \bar u  P_X d  \, \bar u \, \sigma^{\mu\nu} \, P_X s  | K^+ \rangle \otimes j^X_{\ell \ell^\prime , \mu\nu}  = 
8 \left \{ \left ( 1 - \frac{1}{4 N_c} \right ) \, 
\langle \pi^-  | \bar u  P_X d | 0 \rangle  \, \langle 0 |  \bar u \, \sigma^{\mu\nu} \,  P_X s  | K^+ \rangle \right . \\
\\
\qquad  \qquad \qquad \qquad  \qquad \qquad 
-  \frac{1}{4 N_c}  \, \langle \pi^-  | \bar u \, \sigma^{\mu \nu} \, P_X d  | 0 \rangle  \, \langle 0 |  \bar u P_X s  | K^+ \rangle \nonumber \\
\\
\qquad  \qquad \qquad \qquad  \qquad \qquad 
-  \left .  \frac{i}{4 N_c}  \, \langle \pi^-  | \bar u \, \sigma^{\mu \rho} \, P_X d  | 0 \rangle  \, \langle 0 |  \bar u \sigma^\nu_\rho \, P_X s  | K^+ \rangle  \right \} \otimes j^X_{\ell \ell^\prime , \mu\nu} 
\, + \, \dots \, , \\
\\
8 \langle \pi^-  | \bar u  P_X d  \, \bar u \, \sigma^{\mu\nu} \, P_Z s  | K^+ \rangle \otimes j^Z_{\ell \ell^\prime , \mu\nu}  = 
8 \left \{ \langle \pi^-  | \bar u  P_X d | 0 \rangle  \, \langle 0 |  \bar u \, \sigma^{\mu\nu} \,  P_Z s  | K^+ \rangle \right . \\
\\
\qquad  \qquad \qquad \qquad  \qquad \qquad 
-  \left .  \frac{i}{N_c}  \, \langle \pi^-  | \bar u \, \gamma^\mu \, P_X d  | 0 \rangle  \, \langle 0 |  \bar u \gamma^\nu \, P_Z s  | K^+ \rangle  \right \} 
\otimes j^Z_{\ell \ell^\prime , \mu\nu} 
\, + \, \dots \, . \\
\end{array}
\end{equation}

\item {\bf ${\cal O}_7^{\prime \, XZZ}$} \\

\begin{equation}
\label{eq:fierzconnectedO7XZZ2}
\begin{array}{l}
8 \langle \pi^-  | \bar u  \sigma^{\mu\nu} \, P_X d  \, \bar u \, P_X s  | K^+ \rangle \otimes j^X_{\ell \ell^\prime , \mu\nu}  = 
8 \left \{ \left ( 1 - \frac{1}{4 N_c} \right ) \, 
\langle \pi^-  | \bar u  \sigma^{\mu\nu} \,  P_X d | 0 \rangle  \, \langle 0 |  \bar u \, P_X s  | K^+ \rangle \right . \\
\\
\qquad  \qquad \qquad \qquad  \qquad \qquad 
-  \frac{1}{4 N_c}  \, \langle \pi^-  | \bar u \,  P_X d  | 0 \rangle  \, \langle 0 |  \bar u \sigma^{\mu \nu} \, P_X s  | K^+ \rangle \nonumber \\
\\
\qquad  \qquad \qquad \qquad  \qquad \qquad 
+  \left .  \frac{i}{4 N_c}  \, \langle \pi^-  | \bar u \, \sigma^\mu_\rho \, P_X d  | 0 \rangle  \, \langle 0 |  \bar u \sigma^{\nu \rho} \,  P_X s  | K^+ \rangle  \right \} 
\otimes j^X_{\ell \ell^\prime , \mu\nu} 
\, + \, \dots \, , \\
\\
8 \langle \pi^-  | \bar u  \sigma^{\mu\nu} \, P_Z d  \, \bar u \, P_X s  | K^+ \rangle \otimes j^Z_{\ell \ell^\prime , \mu\nu}  = 
8 \left \{ \langle \pi^-  | \bar u  \sigma^{\mu\nu} \,  P_Z d | 0 \rangle  \, \langle 0 |  \bar u \, P_X s  | K^+ \rangle \right . \\
\\
\qquad  \qquad \qquad \qquad  \qquad \qquad 
+  \left .  \frac{i}{N_c}  \, \langle \pi^-  | \bar u \, \gamma^\mu \,  P_Z d  | 0 \rangle  \, \langle 0 |  \bar u \gamma^\nu \,P_X s  | K^+ \rangle  \right \} 
\otimes j^Z_{\ell \ell^\prime , \mu\nu} 
\, + \, \dots \, . \\
\end{array}
\end{equation}

\item {\bf ${\cal O}_8^{XXX}$} \\
\begin{equation}
\label{eq:fierzconnectedO8XXX}
\begin{array}{l}
8 \langle \pi^-  | \bar u  \sigma^{\mu\rho} \, P_X d  \, \bar u \, \sigma^\nu_\rho \, P_X s  | K^+ \rangle \otimes j^X_{\ell \ell^\prime , \mu\nu}  = 
8 \left \{ 
\left(1-\frac{1}{16N_c}\right)\langle \pi^-  | \bar u  \sigma^{\mu\rho} \,  P_X d | 0 \rangle  \, \langle 0 |  \bar u \sigma^\nu_\rho \, P_X s  | K^+ \rangle \right . \\
\\
\qquad  \qquad \qquad \qquad  \qquad \qquad 
+ \frac{i}{2 N_c}  \, \langle \pi^-  | \bar u \,  P_X d  | 0 \rangle  \, \langle 0 |  \bar u \sigma^{\mu \nu} \, P_X s  | K^+ \rangle \nonumber \\
\\
\qquad  \qquad \qquad \qquad  \qquad \qquad 
-  \left .  \frac{i}{2 N_c}  \, \langle \pi^-  | \bar u \, \sigma^{\mu \nu} \, P_X d  | 0 \rangle  \, \langle 0 |  \bar u \, P_X s  | K^+ \rangle  \right \} 
\otimes j^X_{\ell \ell^\prime , \mu\nu} 
\, + \, \dots \, , \\
\end{array}
\end{equation}
\end{enumerate}

\section{{\cal B}-parameters for $K^+ \to \pi^- \ell \ell$ transitions at large $N_c$}
\label{app:Bparameters}

Without a non-perturbative computation of the hadronic matrix elements
$\langle \pi^- | {\cal O}_n^{XYZ}(\mu) | K^+ \rangle$, the best we can do is what is shown in the fourth column of Tabs.~\ref{tab:results1}, \ref{tab:results2} and \ref{tab:results3}. We will, however, sketch 
hereafter the procedure that should be carried on in order to 
go one step further and match the present precision in the computation 
of Wilson coefficients of $\Delta F = 2$ Lagrangians (with $F$ a flavour-related quantum number, such as the strangeness $S$, \cite{ETM:2010ubf,Boyle:2012qb,Boyle:2017skn}). 
For definiteness, we consider a would-be lattice computation of 
the hadronic matrix elements. Our procedure, however, does not depend
on the non-perturbative method adopted and  is completely general. 

\subsection{The standard definition of the $B$-parameters}
\label{sec:Bparameters1}

In eq.~(\ref{eq:mesondecaywidthoneloop}) we introduced the meson decay width at the scale $\mu$, summing over all the spin-averaged amplitudes
$\bar {\cal A}_n^{XYZ}$ defined in eq.~(\ref{eq:renormalizedamplitudes}). The hadronic amplitudes are, on the other hand, computed non-perturbatively as follows: 
\begin{equation}
\label{eq:standardBparameter}
{\cal A}_n^{XYZ}(\mu) = 
\left (
\frac{\langle \pi^- | {\cal O}_n^{XYZ}(\mu) | K^+ \rangle}{
\left [{\cal A}_n^{XYZ}\right ]_{\rm VIA}
}
\right ) \, \left [{\cal A}_n^{XYZ}\right ]_{\rm VIA} = 
{\cal B}_n^{XYZ} (\mu) \, \left [{\cal A}_n^{XYZ}\right ]_{\rm VIA} \, .
\end{equation}
The $B$-parameter ${\cal B}_n^{XYZ}(\mu)$ represents, therefore, 
the non-perturbative correction to the VIA that we are using 
to estimate the hadronic matrix elements. From a numerical point of view, it is much better to compute non-perturbatively the ratio
of the three-point correlation function in the numerator
normalized with the product of two two-point correlation functions
in the denominator, as in the ratio the statistical fluctuations
over the gauge configurations wash out and we may get a more stable 
signal. Some systematics cancel in the ratio, too. For this
reason, numerical computation on the lattice of hadronic matrix elements is usually given in terms of $B$-parameters and not of the matrix elements themselves. 

\subsection{Going from $\mu$ to the NP scale}
\label{sec:NPcomputation}

As we have stressed above, the matrix elements of these operators between initial and final meson states should be computed non-perturbatively at the scale $\mu$. However, quite in general, these matrix elements are computed at yet another scale, 
$\Lambda_{\rm NP}$. In the case of a lattice QCD computation, the scale at which the non-perturbative computation takes place is $a^{-1}$, where the lattice spacing is fixed by measuring
some external quantity (such as the mass of the pion on that lattice, for example) with, in general, $a^{-1} < \mu$. We have, then, to run upward the results using RGE from $a^{-1}$ up to $\mu$. This is done as follows: 
\begin{equation}
{\cal A}_n^{XYZ}(\mu) = 
Z_{nm} (\mu, a^{-1}) \, {\cal B}_m^{XYZ} (a^{-1}) \, 
\left [{\cal A}_m^{XYZ}\right ]_{\rm VIA}
\, ,
\end{equation}
where the $B$-parameter in the rhs takes into account the exchange of soft, long-distance, ${\cal O}(\Lambda_{\rm QCD})$ gluons and
the renormalization constant matrix $\hat Z$ is computed in some convenient regularization and renormalization scheme. If, for 
example, the hadronic matrix element is computed on the
lattice, $\hat Z$ must be computed either perturbatively through lattice perturbation theory \cite{Lepage:1992xa,Capitani:2002mp} or
non-perturbatively in the RI-MOM scheme \cite{Martinelli:1994ty,Donini:1995xj, Donini:1999sf}) on the lattice. In order to remove the scheme dependence introduced by the mismatch between Wilson coefficient and hadronic matrix element regularization and renormalization, we should still (perturbatively) match the non-perturbative matrix element and the Wilson coefficient. This is done through yet another matching matrix $\hat R$, such that:
\begin{equation}
\label{eq:renormalizedlagrangian}
{\cal L}_{\rm eff}^{\Delta L = 2} = \frac{1}{\Lambda_{\rm EW}^5} \, 
\sum_i C_i (\mu) R^{{\rm lattice} \to \overline{\mathrm{MS}} }_{ij} (\mu) \, 
Z_{jk} (\mu, a^{-1}) \, {\cal B}_k^{XYZ} (a^{-1}) \, 
\left [{\cal A}_k^{XYZ}\right ]_{\rm VIA} \, ,
\end{equation}
where $\hat R^{{\rm lattice} \to \overline{\mathrm{MS}} } (\mu)$ is the matching matrix that relates the matrix elements computed 
in a given lattice regularization and renormalization scheme
(here summarized as "lattice") with
the Wilson coefficients computed in dimensional regularization in the
$\overline{\mathrm{MS}}$ scheme, at the scale $\mu$. After performing this procedure, the resulting effective Lagrangian is $\mu$-independent
up to higher-orders in $\alpha_s$.

In this paper, in the absence of a non-perturbative computation of the
hadronic matrix element, we have made the following approximation: 
\begin{equation}
{\cal A}_n^{XYZ}(\mu) \equiv \left [{\cal A}_n^{XYZ}\right ]_{\rm VIA}
\end{equation}
that, after the inclusion of the connected diagrams, represent
indeed a consistent non-perturbative computation up to ${\cal O}(1/N_c)$
in the $1/N_c$ expansion. However, this approximation implicitly
implies that: 
\begin{equation}
\left . \vec{{\cal B}}^{XYZ} (\mu) \right |_{\overline{\mathrm{MS}} }
= \hat R^{{\rm lattice} \to \overline{\mathrm{MS}} } \, \hat Z (\mu, a^{-1}) \, 
\vec{{\cal B}}^{XYZ} (a^{-1}) = \bf I \, + \dots \, .
\end{equation}
This is not always the case. For example, let's think about
the matrix element of the operator ${\cal O}_8^{XXX} (\mu)$. 
In the VIA, this matrix element vanishes, as it is proportional 
to the two-point correlation function of the tensor current
between the vacuum and one meson state. This means that we have
no (non-perturbative) estimate of the value of this matrix element, 
as our only guideline (the VIA) states that this matrix element
is zero. For this reason, we are not able to constrain the corresponding
Wilson coefficient, that is therefore unbounded.

\subsection{A better definition of the $B$-parameters}
\label{sec:Bparameters2}

We have introduced in Sect.~\ref{sec:Bparameters1} the standard
definition of the $B$-parameters used to compute the non-perturbative
contribution of long-range ${\cal O}(\Lambda_{\rm QCD})$ 
soft gluons to the hadronic matrix element at hand. However, 
as stressed above, this definition is unsuitable in the case
the VIA estimate of the matrix element vanishes, since the denominator
of the observable to be computed is ill-defined. This is not
the only problem of the standard definition of the $B$-parameters, though. Consider, for example, the amplitudes corresponding 
to the matrix elements of the operators ${\cal O}_1^{XYZ}$
and ${\cal O}_3^{XYZ}$ for $X \neq Y$. In both cases, the
VIA estimate of the amplitude, that includes ${\cal O}(1/N_c)$ 
corrections due to the connected diagrams, mixes between themselves
the disconnected contributions of the two operators that, in the
VIA, are proportional to the product of two pseudo-scalar densities
or two axial currents, respectively. Now, as we stressed in Sect.~\ref{sec:VIA}, the meson decay constants do not run 
(as their renormalization are finite, due to the PCAC), whereas
the pseudo-scalar densities do (as they depend on the quark running masses, computed in a given renormalization scheme at a given renormalization scale). Also in this case, therefore, the denominator
of the $B$-parameter is not optimal, as it makes more complicated
the renormalization of the $B$-parameter. In Ref.~\cite{Conti:1997gj} it was proposed to normalize operators whose VIA mixed two-point correlation functions of axial currents and pseudo-scalar densities with the corresponding disconnected diagrams, only: 
\begin{equation}
\label{eq:standardBparameter}
{\cal A}_n^{XYZ}(\mu) = 
\left (
\frac{\langle \pi^- | {\cal O}_n^{XYZ}(\mu) | K^+ \rangle}{
\left [{\cal A}_n^{XYZ}\right ]_d
}
\right ) \, \left [{\cal A}_n^{XYZ}\right ]_d = 
\tilde {\cal B}_n^{XYZ} (\mu) \, \left [{\cal A}_n^{XYZ}\right ]_d \, .
\end{equation}
The RGE for the $B$-parameter defined as above is
\begin{equation}
\mu \, \frac{d \tilde {\cal B}_n^{XYZ} (\mu)}{d \mu} = 
\left (  \gamma_{{\cal O}_n} - 2 \gamma_{\cal P} \right ) \, 
\tilde {\cal B}_n^{XYZ} (\mu) \, ,
\end{equation}
where $\gamma_{{\cal O}_n}$ is the anomalous dimension of the operator
${\cal O}_n$ and $\gamma_{\cal P}$ is that of the pseudo-scalar density. Since the disconnected diagram in the VIA renormalize
with the anomalous dimension $2 \gamma_{\cal P}$ (as only the quark masses run with the scale), the renormalized
$B$-parameter times the amplitude of the disconnected diagram in the
VIA renormalize with the anomalous dimension of the operator, only, 
thus reducing the systematic errors. 

This definition of the $B$-parameter, though, does not resolve
the problem of those matrix elements of operators whose disconnected
diagram in the VIA vanish, such as ${\cal O}_8$. For this reason, 
therefore, it is better to normalize the three-point correlation function with the two-point correlation function of the axial current
that, as we already stressed, gets only a finite renormalization constant on the lattice: 
\begin{equation}
\label{eq:standardBparameter}
{\cal A}_n^{XYZ}(\mu) = 
\left (
\frac{\langle \pi^- | {\cal O}_n^{XYZ}(\mu) | K^+ \rangle}{
\left [{\cal A}_3^{XYZ}\right ]_d
}
\right ) \, \left [{\cal A}_3^{XYZ}\right ]_d = 
\bar {\cal B}_n^{XYZ} (\mu) \, \left [{\cal A}_3^{XYZ}\right ]_d \, .
\end{equation}
This proposal was advanced for the first time in Ref.~\cite{Donini:1999nn} in order to simplify the matching 
between the non-perturbative computation 
of the hadronic matrix elements of $\Delta S = 2$ operators
on the lattice with the $\overline{\mathrm{MS}}$ perturbative computation of the corresponding Wilson coefficients. In our case, 
however, it is even more suited, as it permits a proper normalization
of matrix elements of operators whose estimate in the
VIA would vanish, whilst retaining the useful reduction of statistical 
and systematic errors in the ratio of three- to two-point functions. 

\end{appendices}

\bibliographystyle{unsrt} 
\bibliography{bibdl2meson}

\end{document}